\def\NN{\nonumber}
\def\DS{\displaystyle }
\def\r{\right}
\def\l{\left}

\def\d#1{{\partial \over \partial #1}}

\def\Bra#1{\l\langle #1 |}

\def\Ket#1{| #1 \r\rangle}
\def\sca#1#2{{\l\langle #1 | #2 \r\rangle}}
\def\moy#1{{\l\langle #1 \r\rangle}}
\def\vec#1{{\bf #1}}

\def\be{{\beta}}
\def\ga{{\gamma}}
\def\de{{\delta}}
\def\e{\,{\rm e}}
\def\ps{{\psi}}

\def\vt{{\vartheta}}

\def\si{{\sigma}}
\def\ta{{\tau}}

\def\w{{i\omega}}\def\wp{({i\omega})}
\def\Ga{{\Gamma}}
\def\om{{\omega}}
\def\x{\,x}
\def\xb{\,{\bar x}}

\def\up{{\uparrow}}
\def\dow{{\downarrow}}
\def\nf{{\rm n_F}}

\documentstyle[aps,prb]{revtex}
\input epsf.tex
\begin{document}
\draft
\title{Strong-Coupling Perturbation Theory of the Hubbard Model}
\author{St\'ephane Pairault, David S\'en\'echal, and A.-M. S. Tremblay$^\dagger$} 
\address{Centre de recherche en physique du solide
et D\'epartement de physique}
\address{$^\dagger$Institut canadien de recherches avanc\'ees}
\address{Universit\'e de Sherbrooke, Sherbrooke, Qu\'ebec, Canada J1K 2R1.}
\date{May 1999}
\maketitle
\begin{abstract}
The strong-coupling perturbation theory of the Hubbard model is presented and
carried out to order $(t/U)^5$ for the one-particle Green function in arbitrary
dimension. The spectral weight $A({\vec k},\om)$ is expressed as a Jacobi
continued fraction and compared with new Monte-Carlo data of the one-dimensional,
half-filled Hubbard model. Different regimes (insulator, conductor and
short-range antiferromagnet) are identified in the temperature--hopping integral
$(T,t)$ plane. This work completes a first paper on the subject (Phys. Rev. Lett.
{\bf 80}, 5389 (1998)) by providing details on diagrammatic rules and
higher-order results. In addition, the non half-filled case, infinite
resummations of diagrams and the double occupancy are discussed. Various tests of
the method are also presented.
\end{abstract}


\section{Introduction}
\label{Intro}

The study of strongly-correlated electrons has become in the last decade one of
the most active fields of condensed matter physics. The electronic properties
of an increasing body of materials cannot be described adequately by Landau's
theory of weakly-interacting quasiparticles (Fermi liquid
theory).\cite{Lan57a,Lan57b} Best known are the high-$T_{\rm c}$
superconductors and organic conductors. In both cases, a strong anisotropy and
a narrow conduction band contribute to make the effects of interactions between
electrons (mainly Coulomb repulsion) dramatic.

From a theoretical point of view, quasi-one dimensional (e.g. organic
conductors) and quasi-two dimensional (e.g. high-$T_{\rm c}$  superconductors)
systems are quite different. In one dimension, it is possible to solve
satisfactorily a great number of models~: lattice Hamiltonians such as the
Hubbard model\cite{Hub63} can be solved exactly by Bethe Ansatz,\cite{Lie68}
whereas nonperturbative results can be obtained for  the
Tomonaga-Luttinger\cite{Tom50,Lut63} and the $g$-ology models from
bosonization\cite{Sol79} and  renormalization-group
techniques.\cite{Bou91,Sha94} Conformal field theory has been applied
as well, in particular to the Hubbard model,\cite{Fra90,Fra91} whose Bethe
Ansatz solution has limited practical utility. A unified phenomenology of the
so-called {\it Luttinger liquids} emerges from these works, whose most striking
feature is probably spin-charge separation.\cite{Sch95}

On the other hand, the above methods and results cannot be generalized to the
case of two dimensions, relevant to the CuO$_2$ planes present in all
high-$T_{\rm c}$ superconductors. With a half-filled conduction band, all the
parent compounds of the cuprates are antiferromagnetic insulators, and the main
challenge is to understand, first towards which kind of metal they evolve upon
doping, and secondly whether superconductivity can occur, without a
phonon-mediated coupling, in the vicinity of this antiferromagnetic phase.
Moreover, the linear temperature dependence of the resistivity\cite{Chu96} is a
strong experimental indication that the normal phase of high-$T_{\rm c}$
superconductors is not a Fermi liquid. 

In this context, the Hubbard model\cite{Hub63}
\begin{equation}
\label{HubMod}
{\cal H}=-t\sum_{\moy{i,j }\si}c_{i\si}^{\dag}c_{j\si}
+U\sum_i c_{i\up}^{\dag}c_{i\dow}^{\dag}c_{i\dow}c_{i\up}
\end{equation}
has spurred renewed interest for its ability to account for antiferromagnetic
correlations and the Mott metal-insulator transition. It is also the simplest
model of interacting electrons. In Eq.~(\ref{HubMod}), $t$
represents the hopping amplitude between two neighboring sites in a
tight-binding approximation, and $U$ the strength of the very effectively
screened (and thus taken to be local) Coulomb repulsion. At half-filling and
for $t\ll U$, the Hubbard model is equivalent to a Heisenberg model with
antiferromagnetic exchange $J=4t^2/U$, and its ground state has long-range
N\'eel order in any dimension $d\geq 2$. A Mott transition can occur towards a
metallic state either upon doping, or by increasing the ratio $t/U$. Thus, the
Hubbard model is the prototype of strongly-correlated electrons systems and has
been intensely studied, although a more realistic model of the CuO$_2$ planes of
the cuprates should involve several bands.\cite{Geb97}

Solving the Hubbard model is a difficult problem by itself, even in dimension
$d=1$, where the complexity of the Bethe Ansatz solution prevents one from
actually computing most physical quantities. For example, its spectral weight is
known only in the limit $U\to\infty$ and at zero temperature.\cite{Fav97} In
any dimension $d\geq 2$, only approximate\cite{Hub63,Bar76,Col83,Kot86,Vil96B}
or numerical\cite{Bul94,Dag94,Kob98,Mor90,Pre94,Pre95} methods are available.
Only in the limit $d\to \infty$ do important simplifications occur,\cite{Met89} 
allowing to compute most physically interesting quantities in an essentially
exact way.\cite{Pru93,Geo96} In particular, the Mott transition has been
studied in great detail, and is still under discussion.\cite{Log98,Noa99}
Strong coupling perturbation theory, which considers the
interaction term of the Hamiltonian as dominant, and the kinetic term as a
perturbation, has been somewhat neglected so far. Perturbative series for the
thermodynamical potential have been obtained up to order
$t^4$,\cite{Kub80,Pan91,Bar92} but do not yield dynamical correlation functions.
An original method effectively achieving infinite summations of terms, the
so-called Grassmannian Hubbard-Stratonovich transformation, was introduced
independently by C.~Bourbonnais\cite{Bou85} and S.~Sarker,\cite{Sar88} but
contained great difficulties which were overcome only recently\cite{Pairault98}
by the authors. The purpose of the present paper is to develop the ideas of
Ref.~\onlinecite{Pairault98} in greater detail, and to present new results that
have been obtained in the meantime.

Although we are naturally interested in the two-particle correlation 
functions (magnetic susceptibility, conductivity, compressibility) 
and the phase transitions of the Hubbard model, it has proven simpler
to first elucidate one-particle properties.
Furthermore, spin-charge separation in $d=1$ is mostly visible 
in the spectral weight,\cite{Voi94,Voi98} and it is possible to gain 
significant insight into the metal-insulator transition\cite{Geb97} and 
antiferromagnetic correlations from the spectral weight alone.
Thus, we will focus on the one-particle properties of the Hubbard model
throughout most of this paper.
From an experimental point of view, angle-resolved photoemission
spectroscopy (ARPES) is the probe of choice to measure the spectral weight.
Experiments have recently been conducted on quasi one-dimensional 
$\rm SrCuO_2$\cite{Kim96} and $\rm NaV_2O_5$\cite{Kob99}, and on quasi
two-dimensional $\rm SrCuCl_2O_2$\cite{Wel95} antiferromagnetic compounds,
and further analyzed with the help of exact diagonalizations.\cite{Leu97}

In Sect.~\ref{Method}, we present the strong-coupling expansion for dynamical
correlation functions, and provide explicit examples of diagram calculation in
Appendix~\ref{ExemplesDetailles}. The method is applied to the half-filled
Hubbard model in Sect.~\ref{Application}, and the resulting spectral weight and
double-occupation are compared to Monte-Carlo data in Sect.~\ref{Comparisons}.
Partially self-consistent solutions, involving an infinite sum of diagrams, are
investigated in Sect.~\ref{infresum}, and the question of doping addressed in 
section~\ref{dop}. Appendix~\ref{Atomique} provides the atomic one- and 
two-particle correlation functions, necessary to apply the method, and 
appendix~\ref{TestProc} presents a practical test of the method on a toy model.
Higher-order terms of the expansion are given in appendix~\ref{HODS}. 

\section{The Strong-Coupling Expansion}
\label{Method}

In this section we derive the strong-coupling expansion of correlation
functions for a wide class of Hamiltonians. We first specify the form
that the Hamiltonian should have in order for the method to work. Then we
introduce the Grassmannian Hubbard-Stratonovich transformation, and the
diagrammatic perturbation theory itself.

\subsection{The Hamiltonian}

Consider a Hamiltonian ${\cal H}={\cal H}^0+{\cal H}^1$, 
where the unperturbed part ${\cal H}^0$
is diagonal in a certain variable $i$ (for instance a site variable), and 
let us denote collectively by $\si$ all the other variables of the problem
(for instance a spin variable). 
From now on we will call $i$ the ``site variable'' and $\si$ the 
``spin variable'' for definiteness, though they may represent any set 
of quantum numbers.
This Hamiltonian describes the behavior of fermions, and we suppose that it 
is normal-ordered in terms of the annihilation and creation operators 
$c_{i\si}^{(\dag )}$.
${\cal H}^0$ may be written as a sum over $i$ of on-site Hamiltonians 
involving only the operators $c_{i\si}^{(\dag )}$ at site $i$:
\begin{equation}
{\cal H}^0=\sum_i h_i (c_{i\si}^{\dag},c_{i\si} )\>.
\end{equation}
Whenever doing actual calculations, we will use the Hubbard model.
For the latter, ${\cal H}^0$ corresponds to the atomic limit, that is
\begin{equation}
h_i (c_{i\si}^{\dag},c_{i\si} )=Uc_{i\up}^{\dag}c_{i\dow}^{\dag}
c_{i\dow}c_{i\up}\>.
\end{equation}
We will use the notation $u=U/2$ throughout this paper for convenience.
We suppose that the perturbation ${\cal H}^1$ is a one-body operator:
\begin{equation}
{\cal H}^1=\sum_\si \sum_{ij} V_{ij} c_{i\si}^{\dag}c_{j\si} \>,
\end{equation}
with $V$ a Hermitian matrix. Here, we suppose in addition that the 
perturbation is diagonal in the spin variable, but this needs
not be the case. For the Hubbard model, ${\cal H}^1$ represents the 
kinetic term, and $V$ is the matrix of hopping amplitudes.

Introducing the imaginary-time dependent Grassmann fields $\ga_{i\si}(\ta ),
 \ga_{i\si}^\star(\ta )$, the partition function at some temperature 
$T=1/\be$ may be written using the Feynman path-integral formalism:
\begin{equation}
Z=\int [d\ga^\star d\ga ] \exp-\int_0^\be d\ta \l\{ 
\sum_{i\si}\ga_{i\si}^\star(\ta )\l( \d\ta -\mu \r)
 \ga_{i\si}(\ta ) + \sum_i h_i (\ga_{i\si}^\star(\ta ),
\ga_{i\si}(\ta ) ) + \sum_{ij\si} V_{ij} 
\ga_{i\si}^\star(\ta) \ga_{j\si}(\ta ) \r\} \>.
\end{equation}
In order to lighten the notation, we will use the first few latin letters
($a,b,...$) to denote sets such as $(i,\si,\ta )$, and use bra-ket
notation. For instance:
\begin{equation}
\int_0^\be d\ta \sum_{ij\si} V_{ij} \ga_{i\si}^\star(\ta) 
\ga_{j\si}(\ta )=\sum_{ab} V_{ab} \ga_a^\star\ga_b = \Bra\ga V\Ket\ga \>.
\end{equation}

\subsection{The Grassmannian Hubbard-Stratonovich transformation}
\label{GHST}

At this point, one could use standard perturbation theory and expand the 
$S$-matrix in terms of unperturbed correlation functions. Due to the absence
of Wick theorem in the case of a nonquadratic unperturbed Hamiltonian, this
approach does not lend itself to a satisfactory diagrammatic theory. One
cannot define one-particle irreducibility for the diagrams, and one has to be
very careful in order to avoid double counting of certain contributions. For
example, Pan and Wang,\cite{Pan91} and Bartkowiak and Chao\cite{Bar92} had to
deal with over five hundred different diagrams in order to obtain the
thermodynamic potential of the Hubbard model up to fourth order. These
problems were solved by Metzner,\cite{Met91} who organized the perturbation
series as a cumulant expansion, and formulated diagrammatic rules with
unrestricted sums over momenta. In this subsection we show that Metzner's
results can be obtained in a straightforward fashion and even further
simplified by means of a simple transformation on the partition function. This
transformation was first proposed by Bourbonnais\cite{Bou85} and applied (at
zeroth order) to the Hubbard model by Sarker.\cite{Sar88} D. Boies {\it et
al.} also used it to study the stability of several Luttinger
liquids coupled by an interchain hopping.\cite{Boi95} 

The Grassmannian Hubbard-Stratonovich transformation
amounts to expressing the perturbation part of the exponential 
as the result of a Gaussian integral over auxiliary 
Grassmann fields $\ps_{i\si}(\ta ), \ps_{i\si}^\star(\ta )$ 
as follows:
\begin{equation}
\label{transfo}
\int [d\ps^\star d\ps] \e^{ \Bra\ps V^{-1} \Ket\ps + \sca\ps\ga + 
\sca\ga\ps }=\det(V^{-1})\e^{-\Bra\ga V\Ket\ga} \>,
\end{equation}
In terms of the auxiliary field, the partition function becomes, up
to a normalization factor:
\begin{equation}
\label{z2}
Z=Z_0 \int [d\ps^\star d\ps] \e^{\Bra\ps V^{-1}\Ket\ps}
\moy{\e^{\sca\ps\ga + \sca\ga\ps}}_0 \>,
\end{equation}
where $\moy{...}_0$ means averaging with respect to the 
unperturbed Hamiltonian.
Denoting by $\moy{...}_{0,\rm c}$ the cumulant averages, and owing to the 
block-diagonality of ${\cal H}^0$, the average in Eq.~(\ref{z2})
can be rewritten as:
\begin{equation}
\label{cum}
\exp
\sum_{R=1}^{\infty}{1\over (R!)^2}\sum_{i\{ \si_l,\si '_l\} }
\int_0^\be\prod_{l=1}^R d\ta_l d\ta '_l
\ps_{i \si_1}^\star(\ta_1).. \ps_{i \si_R}^\star(\ta_R)
\ps_{i \si '_R}(\ta '_R)..\ps_{i \si '_1}(\ta '_1)
\moy{ 
\ga_{i \si_1}(\ta_1).. \ga_{i \si_R}(\ta_R)
\ga_{i \si '_R}^\star(\ta '_R)..\ga_{i \si '_1}^\star(\ta '_1)
}_{0,{\rm c}}.
\end{equation}
We will denote by $G^{R({\rm c})}_{a_1..a_R\atop b_1..b_R}=
(-)^R\moy{\ga_{a_1}..\ga_{a_R}\ga_{b_R}^\star..
\ga_{b_1}^\star}_{0,({\rm c})}$ 
the various Green functions of the unperturbed Hamiltonian.
The partition function now takes the familiar form
\begin{equation}
\label{z3}
Z\propto \int [d\ps^\star d\ps] 
\exp\l\{-S_0[\ps^\star,\ps ]-\sum_{R=1}^{\infty}
S_{\rm int}^R[\ps^\star,\ps ]\r\}\>,
\end{equation}
where the action has a free (Gaussian) part
\begin{equation}
\label{s0}
S_0[\ps^\star,\ps ]=-\Bra\ps V^{-1}\Ket\ps\>,
\end{equation}
and an infinite number of interaction terms\cite{note-Boies}
\begin{equation}
\label{sint}
S_{\rm int}^R[\ps^\star,\ps ]={-1\over (R!)^2}
{\sum_{\{a_l,b_l\}}}'
\ps_{a_1}.. \ps_{a_R}
\ps_{b_R}^\star..\ps_{b_1}^\star
G^{R\rm c}_{b_1..b_R\atop a_1..a_R }.
\end{equation}
The primed summation in Eq.~(\ref{sint}) reminds us that all the fields
in a given term of the sum share the same value of the site index.
The free propagator for the auxiliary fermions is just $V$, and we may 
now use Wick's theorem to derive diagrammatic rules in order to treat 
the interaction terms perturbatively.
Appendix~\ref{ExemplesDetailles} gives a thorough description of
the diagrammatic rules as well as two explicit examples of 
application to the one-particle Green function and the thermodynamical
potential.

One may wonder why we did not include the first interaction term into the
free part of the action $S_0[\ps^\star,\ps ]$, since it is quadratic in the
field $\ps$. The reason is that we want to count precisely the order of a
given diagram in the perturbation $V$. By separating the free and interacting
parts of the action as in Eqs.~(\ref{s0},\ref{sint}), the order of a given
diagram is simply the number of $\ps$ propagators, whereas the alternate method
mixes all powers of $V$ from zero to infinity. This question will be discussed
more thoroughly in Sect.~\ref{infresum} below. 
%
\subsection{Electron Green functions}
\label{Relations}

We now show how to deduce the correlation functions of the original electrons
($\ga$) from those of the auxiliary fermions ($\ps$) to which the
perturbation theory applies. When coupling the electron field $\ga$ to an
external Grassmannian source $J_a^\star,J_a$, the partition function takes
the form (very similar to Eq.~(\ref{z2})):
\begin{equation}
\label{zj}
Z(J^\star,J)=\int [d\ps^\star d\ps] \e^{\Bra\ps V^{-1}\Ket\ps}
\moy{\e^{\sca{\ps+J}\ga + \sca\ga{\ps+J}}}_0 \>.
\end{equation}
A $2R$-point correlation function reads
\begin{equation}
\label{corr2}
{\cal G}^{R}_{a_1..a_R\atop b_1..b_R}=
(-)^R \moy{\ga_{a_1}..\ga_{a_R}\ga_{b_R}^\star..\ga_{b_1}^\star}=
{1\over Z}\int [d\ps^\star d\ps] 
\l[
{\de \moy{\e^{\sca{\ps+J}\ga + \sca\ga{\ps+J}}}_0 \over 
\de J^\star_{a_1}..\de J^\star_{a_R}\de J_{b_R}..\de J_{b_1}}
\r]_{J=0\atop J^\star=0} \e^{\Bra\ps V^{-1}\Ket\ps}\>.
\end{equation}
Noticing that
\begin{equation}
\label{deriv}
\l[
{\de \moy{\e^{\sca{\ps+J}\ga + \sca\ga{\ps+J}}}_0 \over 
\de J^\star_{a_1}..\de J^\star_{a_R}\de J_{b_R}..\de J_{b_1}}
\r]_{J=0\atop J^\star=0}=
{\de \moy{\e^{\sca{\ps}\ga + \sca\ga{\ps}}}_0 \over 
\de {\ps}^\star_{a_1}..\de {\ps}^\star_{a_R}\de {\ps}_{b_R}..\de {\ps}_{b_1}}
\>,
\end{equation}
we perform $2R$ integrations by parts, and obtain:
\begin{equation}
\label{part}
{\cal G}^{R}_{a_1..a_R\atop b_1..b_R}=
{(-)^R\over Z}\int [d\ps^\star d\ps]
\l(
{\de \e^{\Bra\ps V^{-1}\Ket\ps} \over \de {\ps}_{b_1}..\de
{\ps}_{b_R}\de {\ps}_{a_R}^\star..\de {\ps}_{a_1}^\star}
\r)
\moy{\e^{\sca{\ps}\ga + \sca\ga{\ps}}}_0 \>,
\end{equation}
which expresses ${\cal G}^R$ in terms of correlation functions of the $\ps$
field. For instance, a straightforward calculation gives the relation between
the Green functions ${\cal G}_{ab}=-\moy{\ga_a\ga_b^\star}$ and
${\cal V}_{ab}=-\moy{\ps_a\ps_b^\star}$, which we write in matrix form
\begin{equation}
\label{gp}
{\cal G}=-V^{-1}+V^{-1}{\cal V}V^{-1}\>.
\end{equation}
If $\Ga$ denotes the self-energy of the auxiliary field,then
\begin{equation}
\label{Green}
{\cal G}=\l( \Ga^{-1}-V \r)^{-1}.
\end{equation}
Likewise, a connected 2-point correlation function of the 
original fermions is the corresponding amputated,
connected $2$-point correlation function of the auxiliary fermions.
\begin{equation}
\label{g2p2}
{\cal G}^{\rm IIc}_{a_1a_2,b_1b_2}=
(V^{-1})_{a_1a_1'}(V^{-1})_{a_2a_2'}
{\cal V}^{\rm IIc}_{a_1'a_2',b_1'b_2'}
(V^{-1})_{b_1'b_1}(V^{-1})_{b_2'b_2},
\end{equation}
where a summation over repeated indices is implicit on the right-hand side.

To conclude this section, let us summarize our line of thought 
up to now. We wish to compute physical quantities within a strongly 
correlated fermion model (say, the Hubbard model). Given that the energy
of the interaction is greater than the bandwidth, we want to build a
strong-coupling expansion. Ordinary perturbation theory turns out to be quite
cumbersome, but the simple transformation (\ref{transfo}) restores Wick's
theorem and a (nearly standard) diagrammatic approach for the auxiliary
fermions. Finally, the simple relations (\ref{gp},\ref{Green},\ref{g2p2})
make the connection back to electron Green functions.
 
\section{Application to the Half-Filled Hubbard Model}
\label{Application}

We apply the method presented in the previous section to the Hubbard model at
half filling in any dimension. When doing so, we first have to deal with an
unexpected causality problem, whose solution can be viewed as part of the
method itself. Then we compute the spectral function and the density of
states, and discuss their relevance to the Mott metal-insulator transition
and to antiferromagnetic correlations.

\subsection{The one-particle Green function up to order three}
\label{FracTrick}

The simplest dynamical quantity amenable to our method is the one-particle
spectral function:
\begin{equation}
\label{A(k,w)}
A(\vec k,\om)=\lim_{\>\eta \to 0^+}
-2\>{\rm Im}\>{\cal G}(\vec k,\om +i\eta),
\end{equation}
It is the simplest energy and wavevector resolved correlation function. As
such, it contains a lot of physical information, far more detailed than the
density of states, which is only its momentum-integrated version. It is also
the best tool to investigate the Fermi liquid character of the system, i.e.,
whether or not it is dominated by a narrow peak at the Fermi level. Finally,
it is very badly known from a theoretical (and numerical) point of view,
whereas experiments of angle-resolved photoemission have recently become more
reliable and precise. To obtain the Green function, we have to compute the
self-energy $\Ga$ of the auxiliary fermions.
From now on we will use the nearest-neighbor Hubbard model
\begin{equation}
{\cal H}=-t\sum_{\moy{i,j }\si}c_{i\si}^{\dag}c_{j\si}
+2u\sum_i c_{i\up}^{\dag}c_{i\dow}^{\dag}c_{i\dow}c_{i\up}
\qquad\l(u\equiv\frac U2\r),
\end{equation}
which means that $V(\vec k)=-2tc(\vec k )$, where
$c(\vec k )=\sum_{m=1}^d\cos k_m$. 
Throughout this section, we work at half-filling
by setting the chemical potential to $\mu=u=U/2$.
The simplest diagram contributing to $\Ga$, of order zero in $t$,
is just the atomic Green function 
\begin{equation}
\Ga^{(0)}(\vec k,\w)=G(\w)={\w\over\wp^2-u^2} \>,
\end{equation}
where $\w$ stand for a complex frequency (we keep the symbol $\om$ for
a real frequency). This leads to the following approximate Green function:
\begin{equation}
\label{rpa}
{\cal G}^{(1)}(\w)={1\over{\DS \wp^2-u^2\over\DS \w}+2t c(\vec k)}.
\end{equation}
One recognizes in Eq.~(\ref{rpa}) 
the result of the Hubbard-I approximation,\cite{} whose properties are well
known. The original atomic levels at $\pm u$ are spread out by the hopping 
term into two symmetric bands called Hubbard bands. The lower band
has band edges at $\pm 2td-\sqrt{(2td)^2+u^2}$ and never reaches the 
Fermi level, however large $t$ may be. Thus, this
approximation fails to describe the Mott metal-insulator transition.
But in the method used here,
the Hubbard-I approximation simply stems from the zeroth order
value of $\Ga$, and we know a systematic way of improving it.

\begin{figure}
\vglue 0.5 truecm
\epsfxsize 6truecm\centerline{\epsfbox{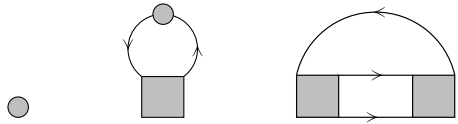}} 
\vglue 0.5 truecm
\caption{
Diagrams contributing to the self-energy $\Ga$ of the auxiliary fermions
up to order $t^3$. 
}
\label{Self3}
\end{figure}

The diagrams contributing to $\Ga$ up to order $t^3$ are presented
in Fig.~\ref{Self3}. 
Actually, there are more diagrams at this order, but they vanish
due to the precise form of $V(\vec k)$. To compute these diagrams,
we need the explicit expression of $G^{\rm IIc}_{\si_1\si_2,\si_3\si_4}
(\w_1,\w_2;\w_3,\w_4)$. The calculation of this atomic correlation
function is a bit cumbersome, though simple in principle.
The main result as well as several useful remarks 
are presented in Appendix~\ref{Atomique}.
For the time being, let us focus on the result:
\begin{equation}
\label{res3}
\Ga^{(3)}(\vec k,\w)={\w\over\wp^2-u^2}+
{6t^2du^2\w\over\l( \wp^2-u^2 \r)^3}+
6t^3 c(\vec k )\l( {{\be u\over 2}
\tanh{\l({\be u\over 2}\r) }\over
\l(\wp^2-u^2\r)^2}+{u^2\l(2\wp^2-u^2\r)\over
\l(\wp^2-u^2\r)^4} 
\r).
\end{equation}
By injecting the above approximate value of $\Ga$ into Eq.~(\ref{Green}) for
the Green function, one obtains a rational function of $\w$, presenting a
finite number of poles in the complex plane. Such a result is already
somewhat disappointing since it cannot account for a continuous spectral weight.
But more seriously, the Green function presents pairs of complex conjugate poles
away from the real axis, and as a consequence does not verify Kramers-Kr\" onig
relations. Furthermore, for a wavevector verifying $V(\vec k)=0$, the Green
function is equal to $\Ga$ itself, and has high-order poles at $\pm u$, leading
to negative spectral weight. So by simply replacing $\Ga$ by its approximate
value to third order in $t$ in Eq.~(\ref{Green}), one ends up with a noncausal
Green function, whose spectral function is neither positive, nor normalized. 

Any Green function acceptable on physical grounds should be causal and have a
positive spectral weight, {\it i.e.}, be a sum (finite or infinite) of simple
real poles with positive residues. We call such a function {\it Lehmann 
representable} since these properties are obvious when considering the Lehmann
representation of the Green function. The way out of the problem is to seek a
Lehmann representable function resembling the approximate expression of $\cal
G$ as much as possible. Since we know $\cal G$ up to third order only, any
function that differs from it by terms of order $t^4$ is {\it a priori} as
good an approximation. A systematic way of building such a function is provided
by a theorem reported in Ref.~\onlinecite{Gil78}: a {\it rational function}
is Lehmann representable {\it if and only if} it can be written as a Jacobi
continued fraction
\begin{equation}
\label{Jacobi}
{\cal G}_J(\w)={a_0\over \w +b_1 -}\>
{a_1\over\w +b_2-}\>...\>
{a_{n-1}\over \w +b_n}\>,
\end{equation}
with 
\begin{equation}
\label{CO}
b_l\> {\rm real\> and}\> a_l>0.
\end{equation}
In any finite system, the exact Green function can be cast into the form of
Eq.~(\ref{Jacobi}) with a finite value of $n$ (the number of {\it floors} of
the continued fraction). If we consider the coefficients $a_l$ and $b_l$ as
functions of
$t$ and expand
${\cal G}_J(\w)$ in powers of $t$, we destroy its continued fraction structure,
and may well obtain an unacceptable approximation. On the other hand, if we
expand the coefficients themselves to some finite order in $t$ and leave them
where they  belong in ${\cal G}_J(\w)$, we can still expect their truncated
Taylor series to verify conditions Eq.~(\ref{CO}), and consequently the
approximate
${\cal G}_J(\w)$ obtained this way to be Lehmann representable. In addition,
their exists a double recurrence relation giving the $a_l$ and $b_l$ starting
from the moments of the function. Since we know the exact Taylor expansion of
$\cal G$ up to order $t^3$, we know all its spectral moments to the same
precision. So all we have to do is to work out the recurrence relations and
compute the coefficients of the continued fraction up to the best precision
available. Once an $a_l$ is found to be zero to the best of our knowledge
--~that is, if it contributes to the full ${\cal G}_J$, given the preceding 
coefficients, at an order higher than the working precision ($t^3$ in this
section)~-- we have to truncate the continued fraction. In all the cases that
we treated, such an $a_l$ always occurred rather quickly (the deepest
continued fraction we had to deal with had eight floors). This means that the
density of poles is always weak, and the poles remain essentially well
separated: they cannot account for the precise shape of a continuous spectral
weight, but can sample it adequately (the moments of the distribution are well
represented). More important, the approximate Green function thus obtained is
causal.

In order to test both the diagrammatic theory and the continued fraction
representation, we treated the exactly solvable case of the atomic limit
itself, away from half-filling. Indeed, setting the chemical potential to $\mu
= u+t_0$, one can either compute the exact Green function, then expand the
solution in powers of $t_0$, or notice that the shift of chemical potential
has the form of a zero-range hopping and obtain the $t_0$ expansion from
diagrams. This procedure is presented in Appendix~\ref{TestProc}, and leads
to the following main conclusions: the diagrams actually give the correct
answer, which however presents the same causality problems as already
mentioned, and the continued fraction representation properly builds back the
poles and the residues of the exact solution. We also tested our approach on the
(again exactly solvable) two-site problem, where the same conclusions
apply. 

The above considerations also shed new light on the usual weak-coupling
perturbation theory. In that case too, one gets multiple poles when truncating
the series for $\cal G$, and the way out of this difficulty is to use Dyson's
equation -- valid because Wick's theorem applies directly to the original
fermions -- and to compute the self-energy. If the self-energy is Lehmann
representable, that is, if it possesses an underlying continued fraction 
structure, then the Green function will inherit this structure from it and will
be Lehmann representable too. We did not
find any definite answer as to why the self-energy itself, as obtained from a 
few diagrams, turns out to be acceptable in general, but there are some
plausible explanations. One is that the unperturbed case already presents a
continuum of levels, instead of only two. So if a double pole appears in the
expression of a diagram, it is likely that its isolated contribution will be
negligible in the thermodynamic limit. Also, the vertices of weak-coupling
theories are often local in time or, if retarded, depend only on two times,
whereas here the vertices have a full dependence on all frequencies
entering them, which is quite peculiar. Finally, nothing proves that a finite-order weak-coupling self-energy is always Lehmann representable, and
counter-examples may well exist. 

Applying the above procedure to the result~(\ref{res3}) yields
the following continued fraction:
\begin{equation}
\label{Jacob3} 
{\cal G}_J (\vec k,\w)={1\over 
\w +2 t c(\vec k) -}\>{u^2\over 
\w-{3\be t^3}\tanh{\l({\be u/ 2}\r) }c(\vec k)/u -}\> 
{6 t^2d\over 
\w -2 t c(\vec k)/d -}\>{u^2\over 
\w +t c(\vec k)/d} 
\>, 
\end{equation}
which has exactly the same Taylor expansion as the exact Green function up
to order $t^3$, verifies the conditions (\ref{CO}), and is normalized to unity.
The next two subsections show that the spectral weight deduced from 
Eq.~(\ref{res3}) describes the Mott metal-insulator transition, and the 
appearance of strong antiferromagnetic correlations at low temperature in the
Hubbard model.

\subsection{The Mott transition}
\label{Mott}

\begin{figure}
\vglue 0.5 truecm
\epsfxsize 8truecm\centerline{\epsfbox{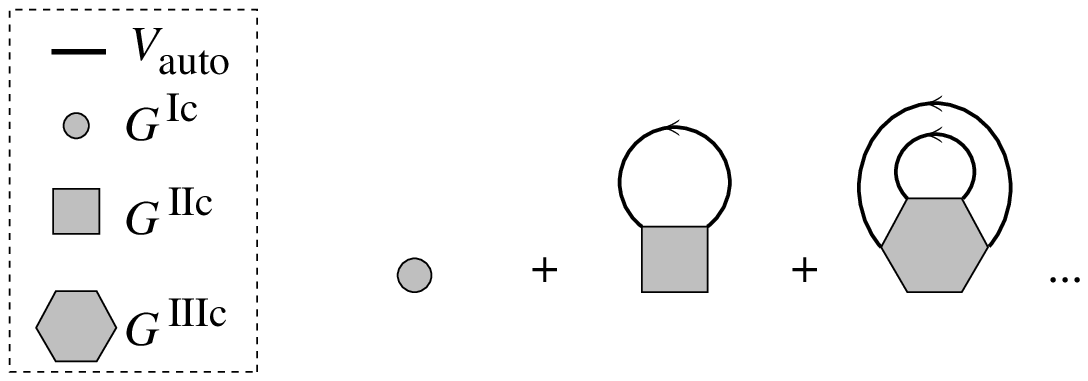}} 
\vglue 0.5 truecm
\caption{Diagrams yielding the exact $\Ga(\w)$ when $d\to\infty$.
$V_{\rm auto}(\vec k,\w)=V(\vec k)/(1-\Ga(\w)V(\vec k))$ is itself
dressed with these diagrams.
}
\label{DiagsdInf}
\end{figure}

Strictly speaking, there is no rigorous definition of the Mott transition
in terms of one-particle properties only. But one can use as a heuristic 
criterion (even at nonzero temperature) the appearance of spectral weight
at the Fermi level. With the normalization (\ref{A(k,w)}) of 
the spectral weight, the density of states has the following expression:
\begin{equation}
N(\om)=\int_{-\pi}^{\pi} {d^d k\over(2\pi)^d}\; A(\vec k,\om) .
\end{equation}
We observe that in the density of states, as $t$ increases
from zero, the two symmetric Hubbard bands -- reduced to two delta functions
at $\pm u$ in the atomic limit -- widen, and eventually mix for 
$t$ beyond some critical value. At high temperature ($T>u$), the gap is
closed by excitations of momentum $\vec k =(0,...,0)$ and 
$\vec k =(\pi,...,\pi)$, making it 
possible to compute the exact value of the critical hopping:
\begin{equation}
\label{tcrit}
t_{\rm c}=u{\sqrt{1+\sqrt{1+12d^2}}\over 2d\sqrt{3}}.
\end{equation}
This gives $U_{\rm c}\simeq 3.2 t$ for $d=1$, to be compared with
$U_{\rm c}\simeq 3.5t$ found in the Hubbard-III approximation. Note that
$t_{\rm c}$ scales properly with the dimension, and leads to $U_{\rm c}=1.86
t^\star$ (with $t^\star=2t\sqrt d$) in the limit of infinite dimension. This
critical value of interaction strength is too large\cite{Geo93,Pru93} for two
reasons. First, when $d\to\infty$ our criterion corresponds to subbands that
meet with an exponentially small density of states, therefore not yet truly
closing the gap. Secondly, when $d\to\infty$, the order of a diagram
has to be counted in a different way. Any nonlocal contribution becomes
negligible, as is visible for example on the contributions of the third-order 
term of Eq.~(\ref{res3}) to the continued fraction Eq.~(\ref{Jacob3}).
Actually, the exact solution in $d\to\infty$ is given by the diagrams
depicted in Fig.~\ref{DiagsdInf}, as was mentioned in
Ref.~\onlinecite{Met91}. Therefore, the approximation~(\ref{res3}) becomes
quite crude in infinite dimension. Note that our criterion for the Mott
transition differs from that used in $d\to\infty$, since we simply demand the
closure of the gap at the Fermi level, without requiring the appearance of a new
coherent peak in $A(\vec k,\om)$ at $\om=0$. Such a peak does not appear in the
present approach.

\begin{figure}
\vglue 0.5 truecm
\epsfxsize 8truecm\centerline{\epsfbox{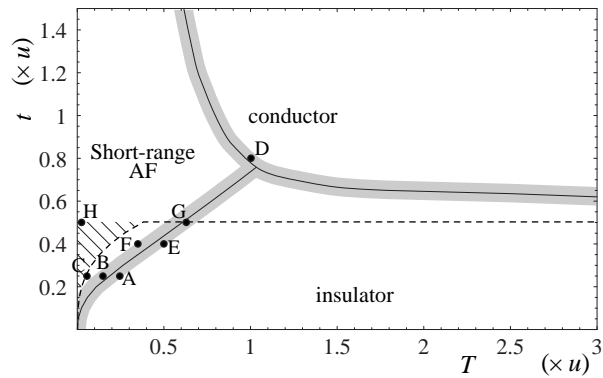}} 
\vglue 0.5 truecm
\caption{
Crossover diagram of the one-dimensional half-filled Hubbard model as obtained
from the third-order Green function.
The estimated domain of validity of the approximation is the region under the
dot-dashed line.}
\label{Crossover}
\end{figure}

When $T<u$, we cannot calculate $t_{\rm c}$ analytically because we do not
know which excitations close the gap, but Fig.~\ref{Crossover} sketches a
numerical evaluation (for $d=1$) in the ($T,t$) plane of the line where the
gap vanishes. The value of $t_{\rm c}$ grows upon lowering $T$, and there is
no Mott transition at zero temperature, in agreement with the exact
result.\cite{Lie68}

One has to keep in mind two important things about the insulator to metal
transition described in this subsection. First, if we call
$\vec Q$ the difference between the momentum of the highest negative energy
(hole-like) excitation and that of the lowest positive energy (particle-like)
excitation, then $\vec Q$ takes the value $(\pi,...,\pi)$ at high temperature,
and goes to zero when $T\to 0$, but the gap always remains indirect. Secondly,
even when the gap has been closed, the spectral weight at the Fermi energy
still has several well separated peaks. $A(\vec k,\om)$ is never dominated at
small $\om$ by a single narrow quasiparticle peak allowing to define a Fermi
wavevector and a Fermi velocity. Thus, when extrapolating it beyond the
transition, our solution does not describe a Fermi liquid. This is not
surprising since we do not expect a perturbative method starting from the
completely localized insulating state to be able to describe both phases
satisfactorily. 

\subsection{Antiferromagnetic correlations}
\label{AFcorr}

\begin{figure}
\vglue 0.5 truecm
\epsfxsize 8truecm\centerline{\epsfbox{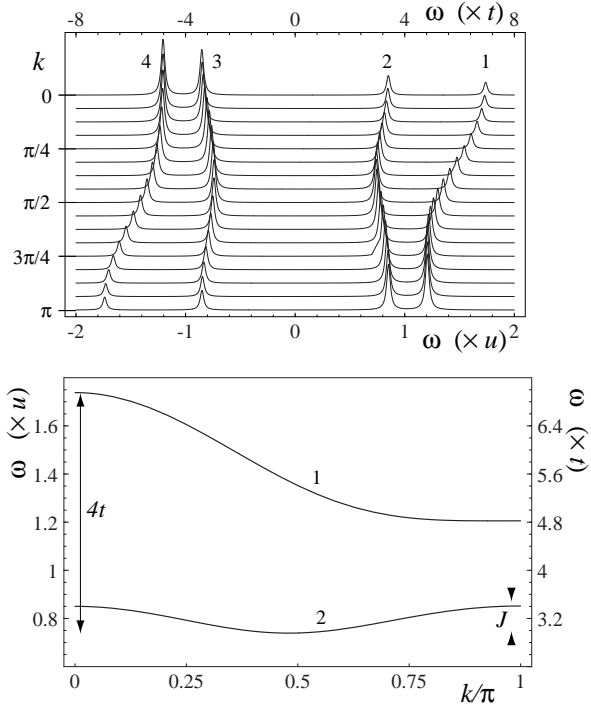}} 
\vglue 0.5 truecm
\caption{
Spectral weight of the half-filled one-dimensional Hubbard model
in the antiferromagnetically correlated regime 
(point C, $t=0.25u$, $T=0.06u=0.24t$).
}
\label{Spec1}
\end{figure}

Here again, a rigorous discussion of the magnetic behavior of the system
should be based on the magnetic susceptibility, which is a two-particle
function. Nonetheless, the spectral weight should contain a valuable
qualitative signature of increased antiferromagnetic correlations, namely a
change in its periodicity. Indeed, if there were a true antiferromagnetic
ordering of the spins, the doubling of the unit cell should lead to a
periodicity of $\pi$ in reciprocal space, instead of $2\pi$, in all the
correlation functions. We expect signs of this reduced periodicity to appear,
even in the disordered phase, when antiferromagnetic correlations become
strong. Their effect actually show up at low $T$, as illustrated by the plot of
$A(k,\om)$ of Fig.~\ref{Spec1}, corresponding to point C of
Fig.~\ref{Crossover}. For simplicity, we discuss the 1D case in this subsection
($\vec k$ becomes $k$), but the conclusions apply to any dimension. We stress
that we only observe the appearance of increased correlations but always remain
in the disordered state: a true transition towards a long-range ordered state
in $d\geq 3$ is not visible in our results, and Fig.~\ref{Crossover} is by no
means a phase diagram, but only a crossover diagram where we distinguished
regions of different qualitative behavior of the one-particle spectral
function. $A(k,\om)$ has four delta peaks (a finite width $\eta $ is added for
clarity) given by dispersion relations $\om_i(k)$, ($i$=1 to 4, as shown in
Fig.~\ref{Spec1}). 

The spectral weight is an even function of $k$, and particle-hole symmetry at
half-filling ensures that $A(k+\pi,-\om)=A(k,\om)$. The minimum of $\om_2(k)$
is at $k=0$ at small $t$ and high $T$ (e.g. point A of Fig.~\ref{Crossover}).
But when $T$ is lowered down to point C (Fig.~\ref{Spec1}), this minimum moves
continuously from $k=0$ towards $k=\pi/2$, and peak 2 loses weight for values
of $k$ much smaller than $\pi/2$. These changes reflect the AF short-range
order that gradually builds up when $T$ becomes smaller than the AF
superexchange $J=2t^2/u$ of the equivalent $t-J$ model. As suggested above, the
approximate cell doubling in direct space translates into a nearly
$\pi$-periodic dispersion for peak 2, although the $2\pi$-periodicity of its
weight and of $\om_1(k)$ reminds us that the state remains paramagnetic. This
is the paramagnetic analog of shadow bands in the spectral weight of the
cuprates. We chose to define an AF crossover line in Fig.~\ref{Crossover} as
the points where $k=0$ ceases to be the minimum of the dispersion $\om_2(k)$.
This crossover line is roughly parabolic ($T\sim t^2$) at low $t$, which
implies a crossover temperature proportional to $J=2t^2/u$.  Furthermore, in
this regime, the width of band 2 is of order $J$ whatever the value of $t$,
supporting the above interpretation. 

\section{Comparison with Numerical Results and Limitations of the Method}
\label{Comparisons}

The purpose of this section is to try and find out where in parameter space --
the $(T,t)$ plane -- the approximate Green function (\ref{Jacob3}) can be
regarded as reliable, to improve it and compare it with Monte-Carlo results
when possible. The entire section applies to the half-filled Hubbard model.

\subsection{Reliability of the third-order solution}
\label{Reliability}

It is not an easy task to define a domain of validity for the present
approximation scheme for several reasons. First, the expansion parameter being
the hopping amplitude $t$, any hopping process is considered on an equal
footing regardless of whether it involves a change in the double occupancy or
not. As a consequence, we expect the ratio $t/T$ to play as important a role as
$t/u$. Secondly the end result involves $t$ in a complicated way, and not only
through a power series. Nevertheless, since the atomic Green function is just 
\begin{equation}
{1\over \DS \w -{\DS u^2\over\DS \w}}
\end{equation}
and gives spectral weight only at $\w=\pm u$, we expect an approximation of
the form~(\ref{Jacobi}) to be valid when the $b_l$'s are small with respect to
$u$. When applied to Eq.~(\ref{Jacob3}), this criterion leads to the
conditions:
\begin{equation}
\label{limites}
2dt<u\quad{\rm and}\quad
\l( {t\over u}\r) ^3<
{1\over 3d}\l({T\over u}\r).
\end{equation}
The first requirement is quite intuitive: it expresses that the bandwidth is
smaller that the on-site interaction, which was the basic assumption anyway.
The second one gives the low-temperature limitation of the method. In dimension
$d=1$, the region where the two conditions Eq.~(\ref{limites}) are fulfilled
lies under the dashed line of Fig.~\ref{Crossover}. When extrapolating the
results outside this region, one cannot predict how fast the approximation
may deteriorate. In particular, it is worth mentioning that the $t\to \infty$
(free-particle) limit is recovered properly. On the other hand, at small $t$
the $T\to 0$ limit again gives a free-particle behavior (except for a singular
behavior for wavevectors such that $V(\vec k)=0$), which is obviously wrong. 
This allows us to define, besides the region of (almost) certain validity
under the dashed line in Fig.~\ref{Crossover}, a hatched region of acknowledged
failure. For the one-dimensional case, the theoretical\cite{Voi98} and
exact\cite{Fav97} results describing spin-charge separation fall in this
region of parameter space, which prevents us from making any meaningful
comparison. However, a definite prediction of our work is that upon raising the
temperature, there appears noticeable spectral weight near $k=\pi$ (for $\om
<0$). This new feature of the spectral weight, visible as peak 3 of
Fig.~\ref{Spec1} and absent from the zero-temperature solutions, might
correspond to the ``question-mark'' features in Fig.~1 of
Ref.~\onlinecite{Kim96}. Thus, temperature seems to have a drastic effect on
the distribution of spectral weight.


\subsection{Beyond third order}
\label{Ordre45}

Requirements such as Eq.~(\ref{limites}) are expected since we are dealing with
a perturbative method. But beside these limitations, our solution 
(Eq.~(\ref{Jacob3})) is plagued with a more serious problem, namely its
discrete spectral weight. Indeed, the exact solution in the thermodynamic
limit certainly has a continuous distribution of poles on the real axis.
Such a continuous distribution is in general necessary to account for
spin-charge separation, or a finite lifetime of one-particle
excitations. Therefore, an approximation describing the spectral weight as a
sum of four delta functions is necessarily a crude one. The
reason why it is difficult to obtain a continuous spectral weight is the huge
degeneracy of the unperturbed Hamiltonian. The starting point being a
collection of independent atoms with the same two energy levels, it is not
likely that a finite-order perturbation scheme can produce a macroscopic
number of distinct approximate eigenvalues. We will discuss in
section~\ref{infresum} a partially self-consistent approach leading to an
extended spectral weight. But in the latter case, we no longer have the freedom
to add higher-order terms in order to solve the causality problem. Facing these
difficulties, we computed the fourth and fifth order of the Green function, in
the hope that they would introduce many more poles into the two narrow Hubbard
bands and give some hints of the lineshape towards which the spectral weight
tends. Let us stress straight away that one is never certain to improve a
perturbative result by adding higher-order terms. Actually, in most of the
cases where perturbation theory is used, the series is asymptotic rather than
convergent, and there is an optimal order -- typically of the order of the
inverse of the expansion parameter -- beyond which the approximation 
deteriorates quickly with each new term.\cite{Neg88} Computing the fourth and
fifth order presented a substantial technical difficulty, since it involved
the function $G^{\rm IIIc}$ having a different expression for each of the 
$5!=120$ possible time orderings (one of the times can be set to zero). We
overcame this difficulty by designing a symbolic manipulation program dedicated
to this problem, taking advantage of the very systematic form of the
expressions in the atomic limit. Let us mention that up to fifth order -- and
that seems to be general -- the even orders lead to a modification of the
partial numerators ($a_l$) only, and the odd orders lead to a modification of
the partial denominators ($b_l$) only, within the continued fraction
(\ref{Jacobi}). Appendix~\ref{HODS} presents the diagrams and the result for
$\Ga(\vec k,\w)$ up to fourth order. $\Ga(\vec k,\w)$ and the corresponding
continued fraction have been computed up to fifth order included, but are too
lengthy to be presented.

At both orders the Green function has eight poles, among which only six have a
significant residue. Thus, there appears only one more pole in each Hubbard
band with respect to order $t^3$. Furthermore, except for the region where the
series certainly converges, the fifth order may give an answer quite 
different from that of the third or fourth order. In particular, the presence
of diverging coefficients makes the self-energy (in the usual sense) go to
zero very fast upon lowering the temperature, and the solution then coincides
with the free-particle limit, completely missing the antiferromagnetic
correlation effect described in Sect.~\ref{AFcorr}. Thus, despite the
important effort invested in computing orders $t^4$ and $t^5$, little progress
has been achieved: the poles still appear as a sparse sampling of the spectral
weight rather than a faithful fit, and the complexity of the solution has
increased up to a point where it is hardly tractable.

In order to test the analytic validity of our results, we computed the exact
Green function up to fifth order for the exactly solvable two-site problem, and
compared with the diagrammatic approach. The two approaches are proven
consistent with each other and the test is summarized in Appendix~\ref{HODS}. 
In the precise case of two sites, the continued fraction has only four poles
instead of eight, as it should. It is interesting to see that in that case too,
the best approximation can be the third, fourth, or fifth order, depending on
the parameters. Most importantly, the two-site problem confirms that dealing
with a small value of $t/u$ is not sufficient to ensure the accuracy of the
approximation. As a matter of fact, we observed empirically that higher orders
improve the solution for $T>t$, and deteriorate it for  $T\ll t$. Explaining
precisely why perturbation theory fails at low temperature appears difficult;
this is probably related to the huge degeneracy of the system in the atomic
limit.

\begin{figure}
\vglue 0.5 truecm
\epsfxsize 16truecm\centerline{\epsfbox{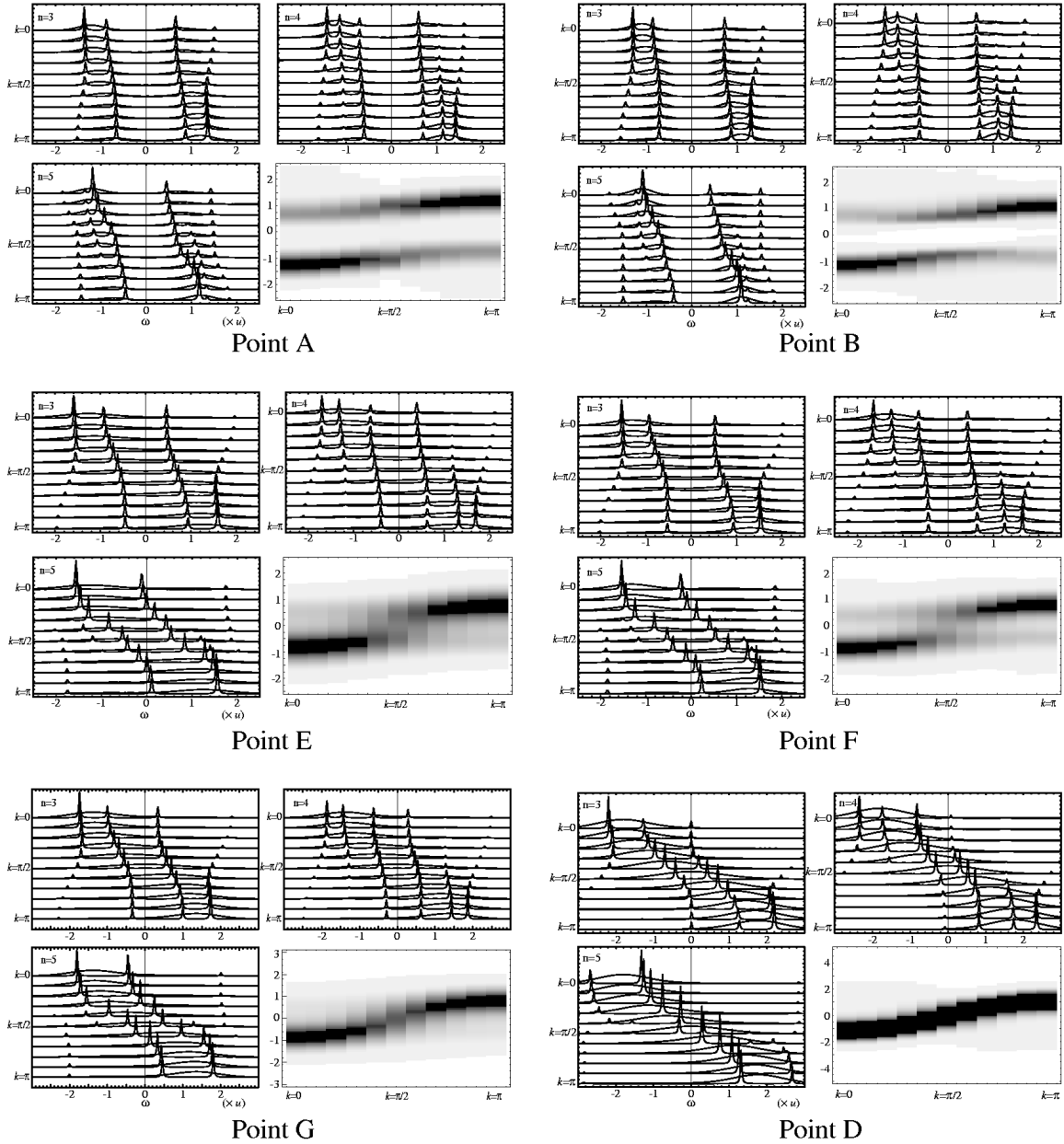}} 
\vglue 0.5 truecm
\caption{
Spectral weight associated with points A,B,E,F,G and D of the crossover diagram
(Fig.~\protect\ref{Crossover}). For each point, we give the spectral weight
obtained from order $t^3$ (top left), $t^4$ (top right), and $t^5$ (bottom
left). A finite width $\eta=0.02$ was added for clarity.
The Monte-Carlo results\protect\cite{Pou98} (smooth curves) were used to
establish the density plot (bottom right). Note that:
\newline Point A lies within the insulating paramagnetic region, as
can be deduced from the presence of a gap and the monotonous dispersion of the
bottom of the upper band (see text for details). For these values of the
parameters, order $t^4$ seems the best approximation.
\newline Point B is in the antiferromagnetically correlated region:
The minimum in the dispersion of the bottom of the band has shifted to $k>0$.
Orders $t^3$ and $t^4$ are best suited there.
\newline Point E is in the insulating paramagnetic region and is best described
by the fourth-order result.
\newline Point F is just at the edge of the antiferromagnetically correlated region.
The third-order approximation is the best one.
\newline Point G is in the metallic region.
Only order $t^5$ gives the closure of the gap for this point.
\newline Point D is well within the metallic region.
}
\label{Spec}
\end{figure}
\subsection{Comparison with Monte-Carlo data}
\label{MCdata}

We now present Monte-Carlo data supporting and illustrating the conclusions
claimed so far, regarding the physical behavior of the model, as well as the 
accuracy of our approximation at various orders in $t$. The
simulations\cite{Pou98} were done for a twenty-site one-dimensional lattice for
reasons of computing power and time. The spectral weight was deduced from
imaginary time Green functions by the Maximum Entropy Method.\cite{Jar96} The
first interesting region is the crossover between the ordinary paramagnetic
insulator and the strongly AF correlated insulator. Fig.~\ref{Spec}
shows the spectral weight for the points A $(t=0.25u,T=0.25)$
and B $(t=0.25u,T=0.125)$ on each side of the crossover. Third-, fourth-, and
fifth-order results are displayed. Of course, A and B being very close to each
other, they have largely similar spectral weights, but their qualitative
difference defined in subsection~\ref{AFcorr} is supported by the density plot.
The fact that orders $t^3$ and $t^4$ are equally well suited for these two
points, located respectively on the lines $T=t$ and $T=t/2$, is consistent
with our argument that Order $t^5$ deteriorates compared to Order $t^3$
precisely in that region. Let us mention that in one dimension, the
Monte-Carlo data show that the antiferromagnetically correlated region is
tiny, and that a spectral function reminiscent of spin-charge separation (with
an important transfer of weight from low to high energy for $k<\pi /2$, $\om
>0$ and $k>\pi /2$, $\om <0$) appears when lowering $T$ a little further
(actually when entering the shaded area of Fig.~\ref{Crossover}). However no
such thing is expected in two dimensions, for which the antiferromagnetically
correlated region extends without spin-charge separation down to $T=0$ where
true long-range order takes place.

The upper border of the paramagnetic insulator region, defined by the closure
of the gap, can be computed more precisely than in Fig.~\ref{Crossover} with 
the help of the fifth order. It leads to a lower value of $t_c$, saturating
around $t_{\rm c}\simeq 0.47u$ at high temperature. Order $t^5$ has been used
when establishing the improved crossover diagram of Fig.~\ref{CrossAm}. The
sequence of points E, G, D, taken along the line $T=1.25t$ for
$t=0.4u,\>0.5u,\>0.8u$, and also depicted in Fig.~\ref{Spec}, shows that the
gap actually closes for a smaller value of $t$ than predicted by order 3. When
lowering the temperature below $t$, the fifth-order result quickly
deteriorates: it does not yield the antiferromagnetic crossover, and predicts
the closure of the gap for smaller and smaller $t_{\rm c}$, with the wrong
conclusion that $t_{\rm c}\to 0$ when
$T\to 0$. Thus, in Fig.~\ref{CrossAm}, we use the third-order result to
obtain the antiferromagnetic crossover line at low temperatures, and we switch
from order $t^3$ to order $t^5$ at the crossing point between the insulator to
metal line (at high $T$) and the antiferromagnetic crossover line (at low $T$).
Therefore, the position of the line separating the paramagnetic metal
(gapless) from an antiferromagnetically correlated insulator (gapped) becomes
conjectural, since fifth order fails to describe it, and third order no longer
brings it to the right ``triple'' point when lowering $t$. Anyway, the density
plot for point F $(t=0.4u,T=0.35)$ (Fig.~\ref{Spec}) clearly shows the
tendency to open a gap upon lowering the temperature.
 
\begin{figure}
\vglue 0.5 truecm
\epsfxsize 8truecm\centerline{\epsfbox{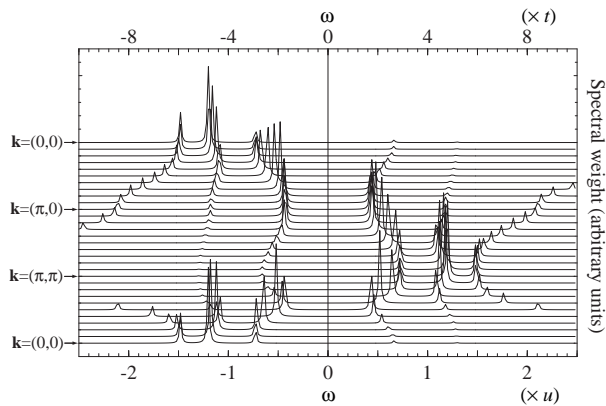}} 
\vglue 0.5 truecm
\caption{
Spectral weight deduced from our approximate Green function at order $t^3$
for Point H2 of Fig.~\protect\ref{CrossAm} ($t=0.25u$, $T=0.025u$) in dimension
$d=2$. This is to be compared with the Monte-Carlo data of
Ref.~\protect\onlinecite{Pre95}. As in dimension $d=1$,\protect\cite{Pairault98}
our result agree quite well with the numerical computations despite the low
value of the temperature.
}
\label{SpecH2}
\end{figure}
\begin{figure}
\vglue 0.5 truecm
\epsfxsize 8truecm\centerline{\epsfbox{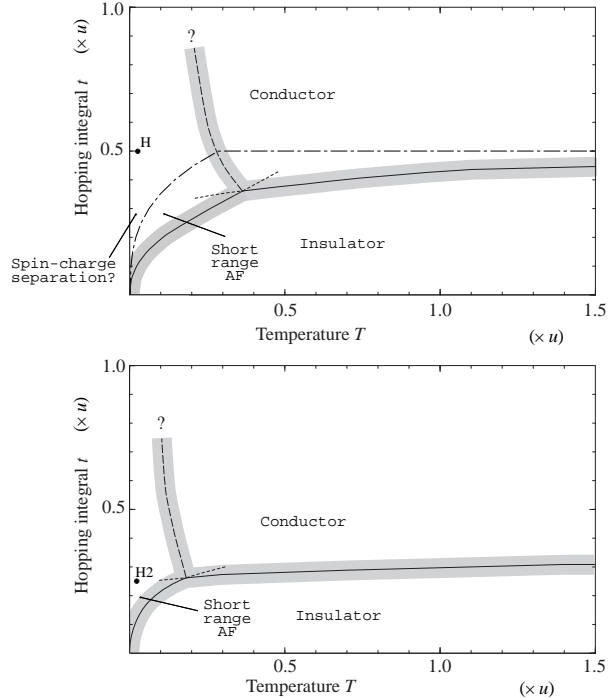}} 
\vglue 0.5 truecm
\caption{
Improved crossover diagram of the half-filled Hubbard model in dimension
$d=1$ (top) and $d=2$ (bottom).
The dot-dashed line reminds the estimated validity region in one dimension.
The antiferromagnetic crossover was calculated with the $t^3$ order result,
and the Mott transition line with the help of order $t^5$.
The limit between the metallic and the insulating antiferromagnetic regions 
is no longer clearly defined. 
}
\label{CrossAm}
\end{figure}
Finally, Fig.~\ref{SpecH2} shows a comparison of the $t^4$ solution and the
Monte-Carlo data of Ref.~\onlinecite{Pre95} for point H2 in the
two-dimensional case. The agreement is excellent, even though the parameters
fall in a region where we do not expect our approximation to be good. Such a
good agreement with Monte-Carlo data outside of the expected region of
validity has also been observed in one dimension.\cite{Pairault98} This
probably means that the criteria (\ref{limites}) are to
stringent. One may instead apply the following {\it a posteriori} criterion:
The approximate spectral distribution is reliable if a large fraction of 
the weight (say, 90\%) falls within $\pm t$ of the original Hubbard bands
at $\om=\pm u$ (this is the case for Fig.~\ref{SpecH2}, as well as for
Fig.~3 of Ref.~\onlinecite{Pairault98}). This softer criterion of
reliability probes the overall effect of hopping on the spectral weight, without
a detailed analysis of partial numerators and denominators, whose effect on
the spectral weight is often far from obvious.

To summarize, the strong-coupling expansion has two serious limitations: (i)
one cannot expect to describe faithfully the lineshape of the spectral function
and (ii) it is impossible to obtain an accurate result for any temperature much
lower than $t$. We add that going beyond fifth order within the
systematic approach described in this paper would be difficult and unpractical,
since the (intractable) result would be relevant only deep in the insulating
region. On the other hand, Monte-Carlo calculations agree well with our results
as far as the overall distribution of the weight is concerned, and confirm the
qualitative conclusions of section~\ref{Application}, summarized in
Fig.~\ref{Crossover}. Thus, we are confident that the improved crossover
diagrams of Fig.~\ref{CrossAm} are reliable.

\subsection{Double occupancy}
\label{doub}

\begin{figure}
\vglue 0.5 truecm
\epsfxsize 8truecm\centerline{\epsfbox{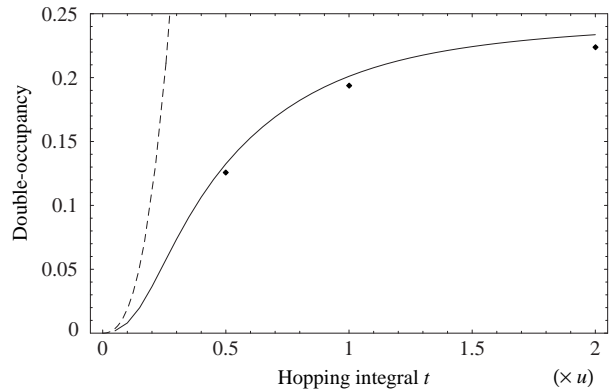}} 
\vglue 0.5 truecm
\caption{
Double occupancy $\moy{n_\up n_\dow}$ as a function of the hopping parameter 
along the line $T=t/6$ in dimension $d=2$.
Comparison between $s_t$ (dashed), $s_J$ (solid), and
Monte-Carlo results\protect\cite{Mor90} (diamonds, bigger than the
uncertainty). Both $s_t$ and $s_J$ are calculated at order $t^4$.  
}
\label{DoubleOccupation_be6}
\end{figure}

\begin{figure}
\vglue 0.5 truecm
\epsfxsize 8truecm\centerline{\epsfbox{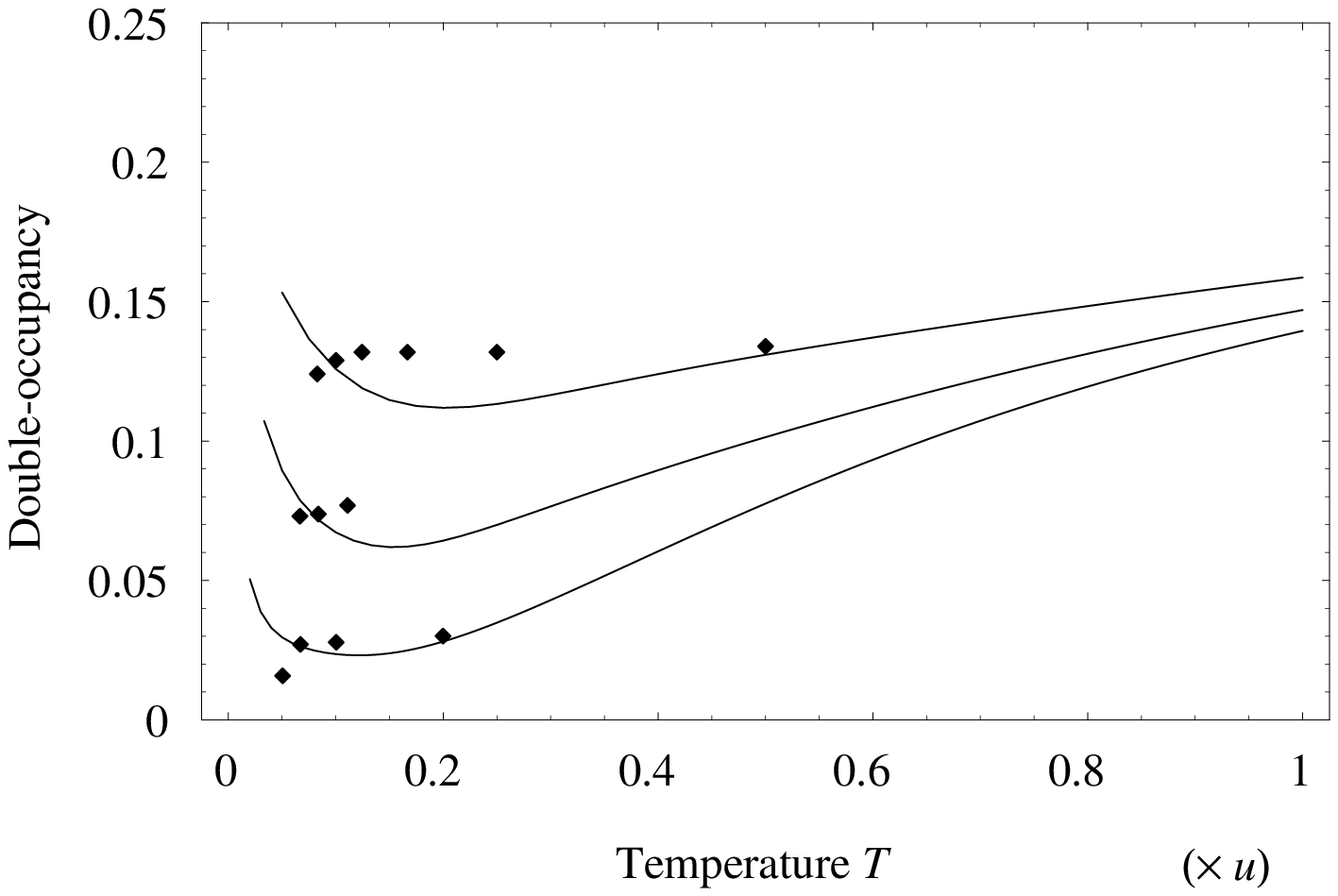}} 
\vglue 0.5 truecm
\caption{
Double occupancy $\moy{n_\up n_\dow}$, computed from $s_J$ at order $t^4$, for
$d=2$  as a function of temperature for various values of $t$ 
($t=0.5u$, $t=0.33u$, $t=0.2u$ from upper to lower curve).
The diamonds represent Monte-Carlo results.\protect\cite{All98} 
}
\label{DoubleOccupation_U4_6_10}
\end{figure}

The double occupancy $\moy{n_\up n_\dow}$ is a local static quantity
essential to the comprehension of the Hubbard model, which gives in
particular the  average potential energy $U\moy{n_\up n_\dow}$.
Except in infinite dimension,\cite{Geo92} $\moy{n_\up n_\dow}$ has to be
determined with the help of Monte-Carlo simulations.\cite{All98,Mor90} It is
possible to deduce it from the one-particle Green function through the exact
relation~:
\begin{equation}
\label{defdoub}
{1\over \be L^d}\sum_{\vec k,\w}
{\cal G}_\si(\vec k,\w)\Sigma_\si(\vec k,\w)  \e^{i0^+}=
U\moy{n_\up n_\dow},
\end{equation}
where $\Sigma_\si(\vec k,\w)$ is the self-energy. There are two possible ways
for us to use the identity~(\ref{defdoub}). On the one hand, from our
knowledge of the thermodynamic potential $\Omega$ (cf. Sect.~\ref{dop} below),
we may extract the power series in $t$ (thereafter denoted $s_t$) of $\moy{n_\up
n_\dow}$: $\beta s_t= -d\Omega/dU$. On the other hand we can insert the
continued fraction (\ref{Jacob3}) into Eq.~(\ref{defdoub}) and compute
$\moy{n_\up n_\dow}$ numerically. This subsection shows that the second option 
(thereafter denoted $s_J$) gives much better results, which confirms that the
continued fraction representation is a controlled way of resumming diagrams. At
half-filling, the power series for the double occupancy is:
\begin{eqnarray}
\label{stserie}
s_t &=&
\frac{1}{1 + \e^\be} + d\l(
\frac{\be^2\e^\be(-1+\e^\be)}{(1+\e^\be)^3} - 
\frac{\be\e^\be}{(1+\e^\be)^2} + 
\frac{-1+\e^\be}{2(1+\e^\be)}\r)t^2
\NN\\
&&+ d\l(
\frac{9( -5 + d )( -1 + \e^\be )}{32(1+\e^\be)} + 
+\be\frac{3(4+7\e^\be-3d\e^\be+4\e^{2\be})}{16(1+\e^\be)^2}
-3\be^2 \frac{( 1 + 7d ) \e^\be ( -1 + \e^\be ) }{16( 1 + \e^\be )^3}\r.\NN\\ 
&&\l.
-\be^3 \frac{3(-1+3d)\e^\be(1-4\e^\be+\e^{2\be})}{8(1+\e^\be)^4}
+\be^4\frac{(-3+7d)\e^\be(-1+\e^\be)(1-10\e^\be + \e^{2\be})}{16(1+\e^\be)^5}
\r){t^4} +{\rm O}(t^6),
\end{eqnarray}
where we have set $u=1$.
Fig.~\ref{DoubleOccupation_be6} sketches $s_t$ and $s_J$ 
as functions of $t$ along the line $T=t/6$ in the two-dimensional case, 
as well as some Monte-Carlo data.\cite{All98,Mor90} 
Whereas $s_t$ quickly increases beyond the maximum physically 
acceptable value of $1/4$, 
$s_J$ interpolates quite well between the low $t$ 
regime and the high $t$ regime where
the form of the continued fraction ensures that the free-particle limit
$\moy{n_\up n_\dow}\to 1/4$ is properly recovered.
The effect of temperature on double occupancy is summarized in 
Fig.~\ref{DoubleOccupation_U4_6_10}, 
showing $s_J$ as a function of $T$ for several values of $t$, as well as
Monte-Carlo results from Refs.~\onlinecite{All98} and \onlinecite{Mor90}. 


$s_J$ always presents a minimum at some temperature, which may be more or less
sharp. Such a minimum is obtained in infinite dimension,\cite{Geo92} where it is
explained by a Pomeranchuk effect. A recent application of a two-particle
self-consistent approach\cite{Dare99} shows that this minimum is still present in
dimension three, but disappears in dimension two: In low dimensions the
self-energy acquires a momentum dependence that makes it more sensitive to
magnetic correlations, thus suppressing the Pomeranchuk effect. In any case, we
believe that this minimum in $s_J$ is an artifact of our approximation, not an
entropy effect. Indeed, when lowering $T$ well below that minimum, we enter the
regime where the self-energy  artificially goes to zero with temperature, and
consequently the double  occupancy goes to $1/4$. We have stressed several times
that this limit is not well rendered by our solution, and it is confirmed by
Monte-Carlo data which show little $T$ dependence at low $T$. Anyway, as long as
we stay in the vicinity of this artificial minimum, our theoretical predictions
differ from the Monte-Carlo values by at most 15\% for $t=0.5u\>(U=4t)$, and the
agreement improves quickly at high temperatures, even for values of $U$ in the
intermediate coupling regime, as in Figs.~\ref{DoubleOccupation_be6} and
\ref{DoubleOccupation_U4_6_10}. It is expected that the agreement will be better
in higher dimension. Moreover, at low temperature, the agreement is better for
smaller values of $t/u$, as expected.

\section{Infinite Resummations}
\label{infresum}

Up to now, we have followed a systematic approach, which consists in building
the series in powers of $t$ of a given quantity. This has led us to interesting
results, but has also shown its limits: (i) it does not yield continuous
spectral functions and (ii) we have pushed the calculations to their maximum
acceptable complexity level. At this point, rather than keeping all diagrams of
a given order, we look for infinite series of diagrams that might embody an
interesting physical feature of the model. 

\begin{figure}
\vglue 0.5 truecm
\epsfxsize 8truecm\centerline{\epsfbox{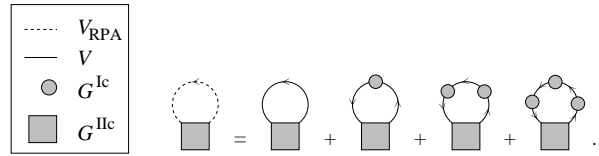}} 
\vglue 0.5 truecm
\caption{
The one-loop ``rosary'' diagram. The simple hopping term
is replaced by an RPA-like propagator.
}
\label{DiagsRPA}
\end{figure}

\begin{figure}
\vglue 0.5 truecm
\epsfxsize 8truecm\centerline{\epsfbox{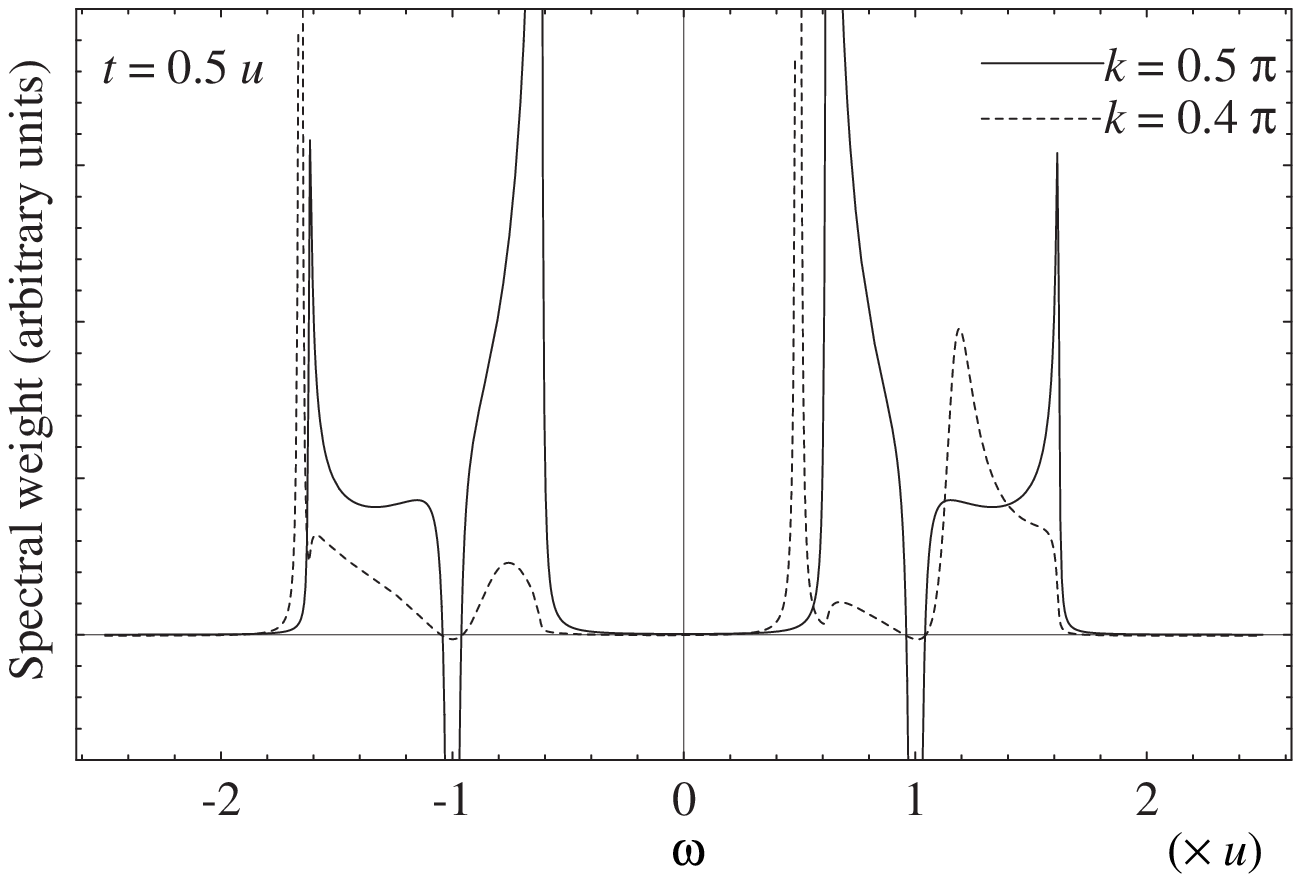}} 
\vglue 0.5 truecm
\caption{
Spectral weight obtained with the auxiliary self-energy of 
Fig.~\protect\ref{DiagsRPA}, with hopping term $t=0.5u$, and momentum
$k=0.5\pi$ (full curve) and $k=0.4\pi$ (dashed curve). The spectral weight is
not everywhere positive within this approximation.}
\label{SpecRPA}
\end{figure}

\subsection{RPA-like propagator}
\label{RPA}

The most natural possibility is to resum all the ``rosary'' diagrams, that 
is, to replace the auxiliary fermion propagator $V(\vec k)$ by
\begin{equation} 
{V(\vec k)\over \DS 1-V(\vec k)G(\w)}
\end{equation}
wherever it appears in the diagrams. This prescription corresponds to an
alternate way of separating free and interacting parts of the action, as
mentioned at the end of Sect.~\ref{GHST}. With this new propagator, the
electron, instead of simply hopping between two neighboring sites, can travel
an arbitrary sequence of one hop (of amplitude
$V(\vec k)$) followed by a period of rest on some site (of amplitude $G(\w)$).
The simplest series of this kind is shown on Fig.~\ref{DiagsRPA}. The result in
$d=1$ is:
\begin{equation}
\label{garpa}
\Gamma (\w)={\w\over\wp^2-u^2}+{3u^2\over \w\l( \wp^2-u^2 \r) }
\l(- 1+{1\over \sqrt {1- \l( {\DS 2t\w \over\DS \wp^2-u^2 }
\r)^2 }}\r).
\end{equation}
Let us comment on $\Ga(\w)$, which coincides with the entire Green function
when $k =\pi/2$. Our main objective has been fulfilled, in the sense that the 
spectral weight $A(\pi/2,\om)$ is now a continuous distribution in the band
\begin{equation}
-t+\sqrt{t^2+u^2}<\om < t+\sqrt{t^2+u^2}
\end{equation} 
and is symmetric. On the other hand, $A(\pi/2,\om)$ 
is not 
everywhere positive. 
Indeed, a detailed analytic study of $\Ga(\w)$ shows that it has 
a positive spectral weight everywhere except at $\om=\pm u$. At those points,
the first term of Eq.~(\ref{garpa}) brings a delta function of weight $1/2$,
and the second term a delta function of weight $-3/2$, resulting in a 
negative peak in the spectral weight. The prefactor $3$ in front of the 
second term has been thoroughly checked, and we have achieved
great confidence that Eq.~(\ref{garpa}) is correct.
When $k\ne \pi/2$, the spectral weight has to be derived from the imaginary
part of 
\begin{equation}
{\cal G}(k,\w)={1\over \Ga(\w)^{-1}+2t\cos (k)}
\end{equation}
when $\w\to\om +i\eta$.
Like for $k=\pi/2$, $A(k,\om)$ is nonzero within the bands, and negative
at $\om=\pm u$. In addition, $A(k,\om)$ has two delta peaks near the edges
of, and outside the bands -- where the Green function has no imaginary part.
Fig.~\ref{SpecRPA} shows $A(0.5\pi,\om)$ 
and $A(0.4\pi,\om)$ for $t=0.5u$.

Thus, the one-loop diagram of leads to a normalized, but nonpositive spectral
weight. We have also lost the interesting physical effects of order three: the
result is temperature-independent and always has a gap at the Fermi level. We
did not find any acceptable way to cure this negative-weight problem. One could
suggest to reduce the normalization of the one-loop diagram of
Fig.~\ref{DiagsRPA}, but introducing such an arbitrary factor is by no means
justified.

\begin{figure}
\vglue 0.5 truecm
\epsfxsize 8truecm\centerline{\epsfbox{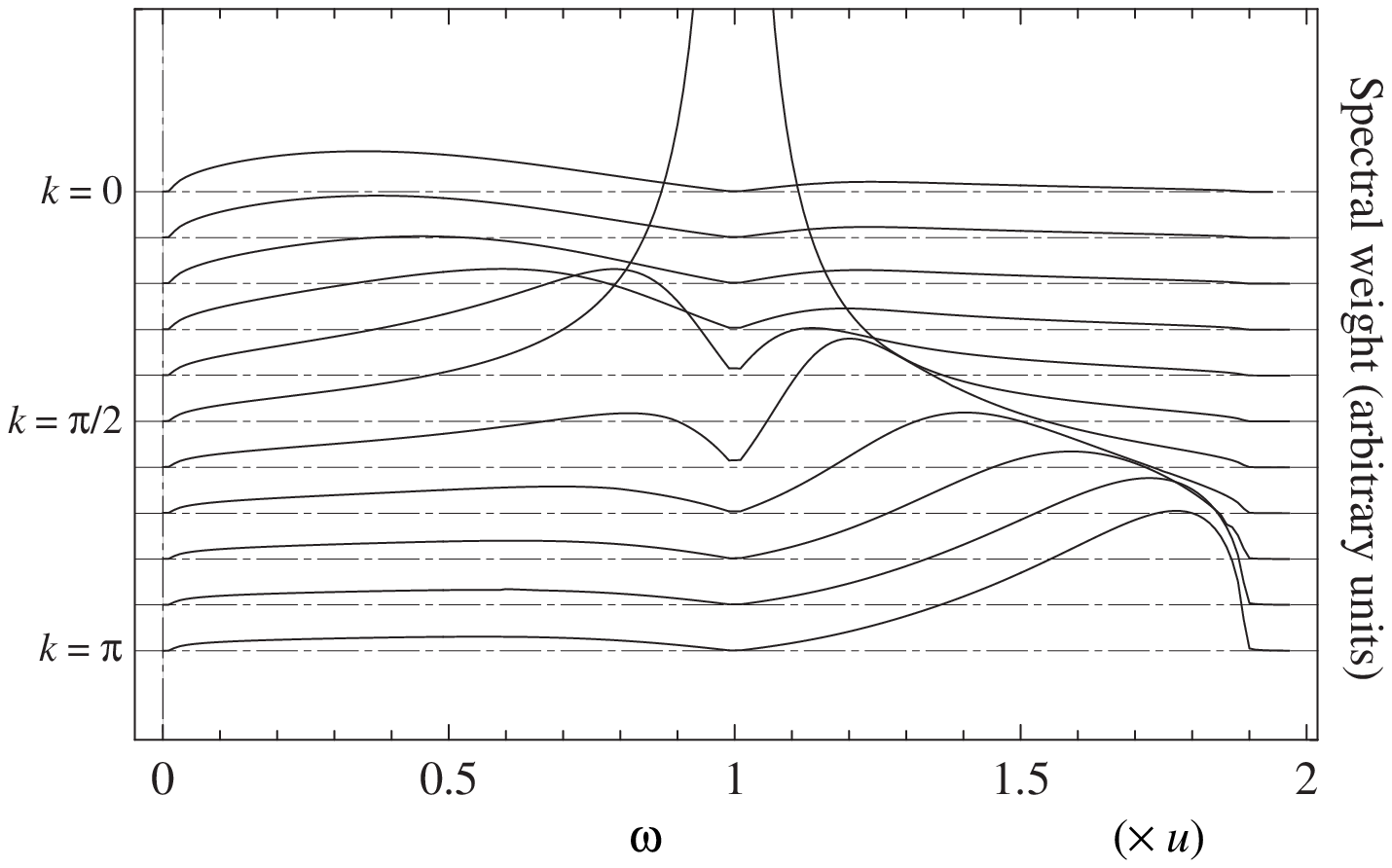}} 
\vglue 0.5 truecm
\caption{
Spectral weight of the one-loop self-consistent solution, for
$t=t_{\rm c}=0.39u$. The wavevector goes from $k=0$ (top) to $k=\pi$
(bottom). The spectral function is clearly not normalized within this
approximation.}
\label{SpecAuto_t039}
\end{figure}

\subsection{One-loop self-consistent approximation}
\label{OLAA}
Another natural attempt involving an infinite subset of diagrams is to compute
the one-loop self-consistent solution: We keep the first two diagrams of
Fig.~\ref{DiagsdInf}, and discard all vertices beyond $G^{\rm IIc}$. The
self-consistent solution $\tilde\Ga(\w)$ obeys the following equation:
\begin{equation}
\label{eqauto}
\tilde\Ga_\si (\w)=G_\si(\w)
-{1\over\be L^d}\sum_{\si_1,\vec k _1, \w_1}
G^{\rm IIc}_{\si,\si_1;\si,\si_1}(\w,\w_1;\w,\w_1)
{V(\vec k _1)\over \DS 1-V(\vec k _1)\tilde\Ga(\w_1)}.
\end{equation}
The sum in Eq.~(\ref{eqauto}) can be computed exactly, given the following
property of $G^{\rm IIc}$ at half filling and zero magnetic field:
\begin{equation}
\sum_{\si_1} G^{\rm IIc}_{\si,\si_1;\si,\si_1}(\w,\w_1;\w,\w_1)
={-3\be u^2\de (\w -\w_1)\over \l( \wp^2 - u^2 \r)^2}
+{\cal E}(\w,\w_1),
\end{equation}
where ${\cal E}$ is an even function of $\w_1$. Since we expect $\tilde\Ga_\si
(\w)$ to be antisymmetric in $\w$, the self-consistent propagator in
Eq.~(\ref{eqauto}) is antisymmetric when $\w_1\to-\w_1$ and $\vec k_1\to
(\pi,...,\pi)-\vec k_1$. Therefore, ${\cal E}$ does not contribute and
\begin{equation}
\label{eqauto2}
\tilde\Ga_\si (\w)=G_\si(\w)
+{3 u^2 \over \l( \wp^2 - u^2 \r)^2}
{1\over L^d}\sum_{\vec k _1}
{V(\vec k _1)\over \DS 1-V(\vec k _1)\tilde\Ga(\w)}.
\end{equation}
Eq.~(\ref{eqauto2}) can easily be solved numerically. Among the various
solutions for the spectral weight, we retain the positive one having a compact
support at very small $t$, and follow its evolution when increasing~$t$. The
result is shown in Fig.~\ref{SpecAuto_t039} for $t=0.25u$, and is obviously not
normalized. The quick suppression of spectral weight close to $\om =\pm u$ when
$k$ differs slightly from $\pi/2$ is a surprising feature too. However, we have 
recovered the closure of the gap for $t\simeq 0.39u$, a value close to the
prediction of Order $t^5$ at high temperature.

The two examples just described show that it is not easy to include 
self-consistency in strong-coupling perturbation theory. 
Lack of positivity or normalization appears in the simplest attempts, and there
is no clue on how to solve this problem. However, it seems the
only way of obtaining a continuous spectral function, and further
developments should probably include a dose of self-consistency.

\section{Doping}
\label{dop}

Studying the Hubbard model at arbitrary filling is important for several
reasons. First of all, the charge gap of the half-filled system disappears upon
doping, and in dimension $d=1$, bosonization shows that this implies dramatic 
changes in the spectral weight. Furthermore, in dimension $d=2$, a slight
doping is directly relevant to the high-T$_{\rm c}$ superconductors. Finally,
the metal-insulator transition can be induced by doping rather than by
interactions. We address the latter aspect in this section. 

On technical grounds, working with arbitrary chemical potential is extremely
difficult, since the complete expression of $G^{\rm IIc}$ given in
appendix~\ref{Atomique} is already hardly tractable, and using it to compute
diagrams would be even worse. Even if it is obviously preferable to include
the full atomic Green function in the unperturbed part, one could treat the
shift in chemical potential $t_0=\mu -u$ as a perturbation, as is done in
appendix~\ref{TestProc}. However, the spatial integrals would no longer cancel
several low-order diagrams involving $G^{\rm IIIc}$ and $G^{\rm IVc}$.
Hence the calculation was carried out with a given $\mu$, without expanding
it as $\mu=u+t_0$, with the help of a suitable generalization of our special
purpose symbolic manipulation program. We were able to compute $\Ga$ up to order
$t^3$ in dimension $d=1$, and to build the corresponding continued fraction. The
partial numerators and denominators of the latter are given in
Eqs.~(\ref{numpart},\ref{denpart}) below.
\begin{eqnarray}
\label{numpart}
a_0&=&1\hfill \NN\\
a_1&=&-4u\big[ -2\be(1-2\nu)^2(\nu^2 - \nu_2)t^2+
(-1+\nu)\nu u+\be^2(1-2\nu)^2
\times(-\nu+4 \nu^2+\nu_2 - 8\nu\nu_2+4\nu_2^2)t^2 u \big] \NN\\
a_2&=&6{t^2}+\frac{4\be\cos(k)(-1+2\nu)(\nu-\nu_2)t^3}{(-1 +\nu)\nu}\NN\\
a_3&=&\frac{-4(-2 - \nu+\nu^2)u^2}{9} - 
\frac{8\be\cos(k)(-1+2\nu)^3(\nu-\nu_2)tu^2}{27(-1+\nu)\nu}
\end{eqnarray}

\begin{eqnarray}
b_1&=&-2t\cos(k)+t_0 -(-1+2\nu)\big[ 4\be(-\nu^2+\nu_2 )t^2+u
+ 2\be^2(-\nu+4\nu^2+\nu_2 - 8\nu \nu_2+4\nu_2^2)t^2 u \big]\NN\\
b_2&=& t_0+(-1+2\nu)u+\frac{2\be(-1+2\nu)t^2}{(-1+\nu)\nu}
\Big\{\nu_2+\nu^4(-2+4\be u )+ \nu^3(2-\be(5+8\nu_2)u)\NN\\
&&- \nu(1+4\be\nu_2^2 u+\nu_2(2+\be u))+ \nu^2(\be u+4\be\nu_2^2 u+\nu_2(2+9\be u))\Big\} \NN\\
&&+\frac{2\be\cos(k)t^3}{(-1+\nu)\nu u} 
\Big\{ \nu^4(-2+4\be u)+ \nu_2(-1+\be\nu_2u)+\nu^2(1+\be u+4\be\nu_2^2 u \NN\\ 
&&+2\nu_2(1+4\be u ))- 2\nu {\nu_2}(1+\be(u+2{\nu_2}u ))- 2{\nu^3}(-1+2\be(u+2{\nu_2}u ))\Big\} \NN\\
b_3&=& \frac{3t_0+u - 2\nu u}{3}+\frac{2\cos(k)t}{9(-1+\nu)\nu}
\Big\{ 2\be\nu_2 u+\nu^2(-9+8\be(1+\nu_2)u)
- 8\be\nu^3 u+\nu(9-2\be(u+4\nu_2 u))\Big\} \NN\\
b_4&=& t_0 +\frac{(-1+2\nu)u}{3}+ \frac{\cos(k)t}{9(-1+\nu)\nu}
\Big\{-4\be\nu_2 u+\nu^2 (9-16\be(1+\nu_2)u)
+16\be \nu^3 u+\nu(-9+4\be(u+4 \nu_2 u))\Big\} \>,
\label{denpart}
\end{eqnarray}
where
\begin{equation}
\nu ={\frac{{\e^{2\be{t_0}}}+{\e^{\be({t_0}+u ) }}}
{1+{\e^{2\be{t_0}}}+2{\e^{\be({t_0}+u ) }}}} 
\end{equation}
is the average number of electrons of a given spin per site in the atomic
limit, and 
\begin{equation}
\nu_2={\frac{{\e^{2\be{t_0}}}}
{1 + {\e^{2\be{t_0}}} + 2{\e^{\be( {t_0} + u ) }}}}
\end{equation}
the double occupancy in the atomic limit. Although complicated,
Eqs~(\ref{numpart},\ref{denpart}) are plausible for several reasons. First,
our symbolic manipulation program was checked on the thermodynamic potential,
and our result agree with Refs.~\onlinecite{Pan91,Bar92}, indicating again
that the result of Ref.~\onlinecite{Kub80} is incorrect. Secondly, the
half-filling case is properly recovered from Eqs.~(\ref{numpart},\ref{denpart}).

At high-enough temperature, the spectral weight derived from 
Eqs.~(\ref{numpart},\ref{denpart}) evolves smoothly with the chemical
potential. For example, for $0<t_0\ll u$, the Fermi level shifts slightly
towards the positive energy peaks, whereas the weight of the latter increases
slightly, with respect to the half-filled case. A similar conclusion applies to
the density of states~: Increasing $\mu$ shifts the Fermi level towards the
upper Hubbard band, and gives the latter more weight, the overall effect being
an increase in the occupation number. In other words, except for a slight
redistribution of weight, the behavior of the system resembles that of a band
insulator.

However, this smooth picture collapses when the temperature is too low,
the spectral weight abruptly becoming negative at various energies, 
which means that the $a_l$'s of Eq.~{\ref{Jacobi}} are no longer all positive.
This is yet another manifestation of the limitation of our method at low
temperature where, according to Refs.~\onlinecite{Pre94,Pre95}, there should be
a quick and massive redistribution of spectral
weight between the Hubbard bands when varying $\mu$. The breakdown of our
solution, which occurs around $T\simeq t$, is concomitant with a
lack of monotonicity in the relation between chemical potential and
filling. This relation is implicitly defined by the thermodynamical potential
through the relation
\begin{equation}
n=-{\partial \Omega\over\partial\mu}\>,
\end{equation}
and becomes ambiguous as soon as the thermodynamical potential is only known
approximately. From this expression a truncated power series in $t$ for $n$
(denoted $n^{(r)}(\mu)$ at order $t^r$) follows directly. Alternately, one may
reverse this relation and express $\mu$ as a truncated power series in $t$,
which we write $\mu^{(r)}(n)$. The two expansions $n^{(r)}(\mu)$ and
$\mu^{(r)}(n)$ may lead to different physical conclusions, which would be
unacceptable. We have verified that, when lowering the temperature, both
functions $n^{(2)}(\mu)$ and
$\mu^{(2)}(n)$, calculated at Order $t^2$, cease to be monotonous nearly at the
same value of $T/t$, which is also the temperature at which the spectral weight
becomes unphysical.

\begin{figure}
\vglue 0.5 truecm
\epsfxsize 8truecm\centerline{\epsfbox{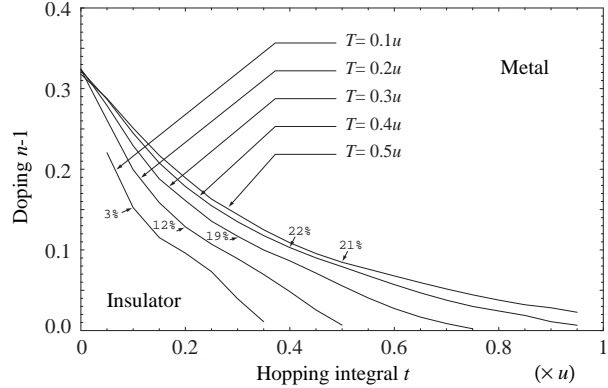}} 
\vglue 0.5 truecm
\caption{
Location of the metal-insulator crossover at various temperatures.
The numbers are estimations of the relative error on $n-1$ at the 
points $t=T$, given that the solution quickly breaks down for $t>T$.
}
\label{MITCroissant}
\end{figure}

Nonetheless, and similarly to what we did in the half-filled case, we can use 
the penetration of the Fermi level into one of the Hubbard bands as a
qualitative criterion for the metal-insulator crossover. This penetration
should translate into an important increase of occupied states (likely to be
delocalized) in either Hubbard band, and therefore an increase in the
conductivity. Fig.~\ref{MITCroissant} shows,  for several temperatures, the
line in the ($t$,$n-1$)-plane where the Fermi level enters the upper Hubbard
band. The numbers associated with each curve are estimations of the accuracy at
the points $T=t$, beyond which the solution breaks down. These numbers are the
relative difference between the results for
$n(\mu)$ obtained in two independent ways~: from the Green function (through
the density of states) and from the thermodynamical potential.

Note from Fig.~\ref{MITCroissant} that, in the atomic limit $t\to0$, the
transition occurs near $n-1=\frac13$. Indeed,
in the atomic limit, it is a simple matter to show that the chemical potential
vanishes at $T=0$ at this filling, and finite temperature corrections are
exponentially small as long as $T\lesssim u$. The chemical potential then
touches one of the (infinitely narrow) atomic bands (at $\om=0$ and $\om=2u$)
and any infinitesimal value of $t$ makes the system conduct.

Thus, in the high-temperature regime where our method is reliable, the 
Mott-Hubbard insulator shares som of the features of a band insulator~: the gap is
nearly rigid, and a crossover towards a good conductivity is obtained when the 
population of carriers ceases to be exponentially small. In contrast to a
band insulator, there is, however, some transfer of spectral weight towards the
band closest to the Fermi level, when
$\mu$ is changed from its half-filled value.

For $T<t$, the expected drastic changes in the spectral weight are not properly
described by our approximate solution, which instead becomes unphysical. 

\section{Conclusion}
\label{conclusion}

In this work, we have developed the ideas of our earlier paper,\cite{Pairault98}
and pushed the computations to higher order. The initial causality problem,
a conspicuous difference between strong- and weak-coupling perturbation theories,
was solved by the continued fraction representation. Among our main results 
are the crossover diagrams of Fig.~\ref{CrossAm}. They demonstrate the
possibility of describing the three qualitatively different regions of parameter
space of the half-filled Hubbard model by a single approximate analytic
expression for the Green function. With respect to Ref.~\onlinecite{Pairault98},
these crossover diagrams were refined by higher-order terms.
Moreover, the crossover diagrams are based on spectral weights that we have shown
to be consistent with Monte-Carlo simulations. The difference between these
spectral weights and their zero-temperature counterparts obtained from other
methods, demonstrates an important temperature dependence of $A(\vec k,\om)$.

We have pointed out the limitations of the simple strong-coupling perturbation
series. Firstly,  our solution breaks down quickly upon lowering the temperature
just after the magnetic correlations have started to show up.  This explains
why the spatial dimension $d$ plays a minor role in our solution.
Actually, as long as magnetic correlations play no role, i.e., at high-enough
temperature, all dimensions are known to show similar behaviors as far as the 
metal-insulator crossover is concerned (we include dimension $d=1$, for which the
gap predicted by Bethe Ansatz is exponentially small in a finite range of $U/t$).
But our method fails at lower temperature, where the system evolves towards
spin-charge separation ($d=1$), or a true antiferromagnetic transition at zero
($d=2$) or finite ($d\geq 3$) temperature.
Secondly, the systematic strong-coupling expansion can only provide
a finite number of poles in the continued fraction and the correponding
approximate Green function cannot have continuous spectral weight. This does not
prevent meaningful comparisons with Monte-Carlo data, since the overall spectral
weight distribution can be assessed. This ``discreteness problem'' might be solved
with the help of a suitable termination function within the continued
fraction,\cite{Vis94}  but this is impossible without any prior  knowledge about
the Green function. The simplest ways of obtaining a continuous spectral weight
from infinite subsets of diagrams lead to nonpositive or unnormalized functions.

Another important difficulty of any expansion about the atomic limit is the
absence of Wick theorem for the atomic limit itself. The auxiliary fermions
allowed us to organize the expansion in the best possible way, and to reduce
tremendously the number of diagrams at a given order. Nevertheless, the exact
$G^{\rm Rc}$'s are still necessary, and the result of  appendix~\ref{Atomique}
for $G^{\rm IIc}$ suggests that they are  quite difficult to compute for $R\geq
3$. Introducing a physically motivated approximation scheme for the atomic
vertices would allow further progress, particularly on the  two-particle
correlation functions. The full two-particle correlation function in the atomic
limit ($U\to\infty$) is a noteworthy byproduct of our work.

The major challenge faced by the strong-coupling perturbation theory is the
low-temperature barrier. We pointed out that the exponentially large degeneracy of
the atomic ground state is likely to be the source of the problem. One possible
solution to this difficulty is to select a ground state more likely to connect
with the low-temperature phases and to organize the perturbation series around
that ground state, which would imply a certain amount of self-consistency. Work
along these lines is in progress.

While this paper focused on properties derived from the one-particle Green
function, two-particle Green functions are also accessible within the method
presented here, but their systematic computation is more involved at order
$t^4$ (it requires the atomic four-particle function $G^{\rm IV}$). For this
reason, we defer discussion of two-particle correlations to a future
publication.

\section{Acknowledgements}

We thank C.~Bourbonnais and N.~Dupuis for many useful discussions. We are
grateful to H. Touchette, S.~Moukouri, L. Chen, and especially D.~Poulin and
S.~Allen for sharing their numerical results. Monte Carlo simulations were
performed in part on an IBM SP2 at the {\it Centre d'applications du calcul
parall\`ele de l'Universit\'e de Sherbrooke}. This work was partially supported by
NSERC (Canada), by FCAR (Qu\'ebec) and by a scholarship from MESR (France) to
S.P.

\appendix

\section{Diagrammatic Rules}
\label{ExemplesDetailles}
This appendix is devoted to deriving and illustrating the diagrammatic  theory
valid for the auxiliary field introduced in Sect.~\ref{GHST}. For definiteness,
we suppose that the site index describes a $d$-dimensional  hypercubic lattice,
and we only consider Hamiltonians where the $h_i$'s do not explicitly depend on
$i$, so that the  interaction terms are translation invariant, in addition to
being  local in space.

The first step consists in expanding the exponential of the interaction terms
of Eq.~(\ref{z3}):
\begin{equation}
\label{z4}
Z=\int [d\ps ^{\star}d\ps] \e^{-S_0[\ps ^{\star},\ps ]}
\sum_{P=0}^\infty {(-)^P\over P!} \l(
\sum_{R=1}^{\infty}S_{\rm int}^R[\ps ^{\star},\ps ] \r) ^P\>.
\end{equation}
For a given $P$, and taking into account the factor ${(-)^P/ P!}$, 
we have a sum of terms of the following form:
\begin{equation}
\label{term1}
{(-)^P\over C_1!..C_H!}S_{\rm int}^{R_1}[\ps ^{\star},\ps ]...
S_{\rm int}^{R_P}[\ps ^{\star},\ps ]\>,
\end{equation}
where $R_1,..,R_P$ are $P$ integers, and $C_1,..,C_H$ the 
multiplicities of the different values that occur in the 
sequence $R_1,..,R_P$.
Suppose we are interested in the following correlation function:
\begin{equation}
\label{corr}
{\cal V}^R_{a_1^0..a_R^0\atop b_1^0..b_R^0}=
(-)^{R_0}\moy{\ps_{a_1^0}..\ps_{a_{R_0}^0}
\ps^{\star}_{b_{R_0}^0}..\ps^{\star}_{b_1^0} }\>.
\end{equation}
The contribution of a given power $P$ to (\ref{corr}) is the sum, for all
possible sequences $R_1,..,R_P$, of
\begin{equation}
\label{term2}
{Z_{\rm Gauss}\over Z}
{\sum_{\{a_r^p,b_r^p\}\atop {p=1..P\atop r=1..R_p}}\kern-0.8em}'
G^{R_1\rm c}_{\{b^1_r\}\{a^1_r\}}..
G^{R_P\rm c}_{\{b^P_r\}\{a^P_r\}}
\moy{
\ps_{a_1^0}..\ps_{a_{R_0}^0}
\ps_{a_1^1}..\ps_{a_{R_1}^1}..
\ps_{a_1^P}..\ps_{a_{R_P}^P}
\ps^{\star}_{b_{R_P}^P}..\ps^{\star}_{b_1^P}
\ps^{\star}_{b_{R_1}^1}..\ps^{\star}_{b_1^1}..
\ps^{\star}_{b_{R_0}^0}..\ps^{\star}_{b_1^0}
}\>_{\atop \rm Gauss},
\end{equation}
times a overall factor 
\begin{equation}
\label{fact}
{(-)^{R_0}\over C_1!..C_H!(R_1!)^2..(R_P!)^2}\>,
\end{equation}
which takes into account the minus signs coming with $S_{\rm int}^R$. Again,
let us stress that  the indices in the sum do not run over all their possible
values,  but rather are restricted to have a common site index if they refer 
to the same vertex (this restriction is encoded in the  primed sum). Since the
action is Gaussian, Wick's theorem is valid, and one can compute 
the average in expression (\ref{term2}) as a sum over all 
possible permutations $\vt$ of $R_0+R_1+...+R_P$ elements:
\begin{equation}
\label{perm}
(\ref{term2})={Z_{\rm Gauss}\over Z} \sum_{\vt}(-)^{\vt}
{\sum_{\{a_r^p,b_r^p\}\atop {p=1..P\atop r=1..R_p}}\kern-0.8em}'
G^{R_1\rm c}_{\{\vt(b^1_r)\}\{a^1_r\}}..
G^{R_P\rm c}_{\{\vt(b^P_r)\}\{a^P_r\}}
\bigg\langle
\ps_{a_1^0}\ps^{\star}_{\vt (b_1^0)}\bigg\rangle_{\rm Gauss}..
\bigg\langle
\ps_{a_{R_P}^P}\ps^{\star}_{\vt (b_{R_P}^P)}\bigg\rangle_{\rm Gauss}
\>,
\end{equation}
where $(-)^{\vt}$ stands for the signature of the permutation, and 
$\big\langle\ps_{a_i^j}\ps^{\star}_{\vt (b_i^j)}\big\rangle_{\rm Gauss}
=-V_{a_i^j \vt(b_i^j)}$. 
A term in the sum over the permutations can be represented easily by 
a Feynman diagram, according to the following rules:

\begin{enumerate}
\item Draw $P$ polygons (vertices) having $2R_1,..,2R_P$ apices 
(internal points) respectively, and $2R_0$ isolated points (external points).

\item To half of the external points and half of 
the apices of each vertex, attach 
an outgoing arrow (a ``bra'', corresponding to a~$\ps$). 
To the rest of the points and apices, 
attach an incoming arrow (a ``ket'', corresponding to~a~$\ps^{\star}$).

\item Label the bras in a conventional order (first the external ones, 
then all the bras of each vertex consecutively), 
and label the kets with corresponding primed integers.

\item To each point assign a latin index, the indices of the 
internal points sharing the same value of the spatial component 
for each vertex.

\item Finish drawing the arrows according to the permutation $\vt$. 
To the lines assign a propagator $V_{ab}$ (bra index first), and 
to the vertices assign a
$G^{R_p\rm c}_{\{b^p_r\}\{a^p_r\}}$ (ket indices first).

\item Integrate over all the internal variables.

\end{enumerate}

\noindent A few remarks are in order. 
First, the ratio $Z_{\rm Gauss}/Z$ allows a restriction to connected diagrams
only.\cite{Abr63} Secondly, it can easily be seen that the exchange of two
equivalent vertices  does not change the value of a diagram, nor does
the exchange of  two internal points of the same type (bra or ket) on the
same vertex.  This allows a restriction to topologically different diagrams,
each of which  with a proper multiplicity factor. 
Thus, to span all the permutations, it is sufficient to consider only the 
topologically distinct diagrams, and to assign them a factor 
\begin{equation}
{(-)^{R_1+..+R_P+\vt}\times{\rm multiplicity}\over C_1!..C_H!(R_1!)^2..(R_P!)^2}.
\end{equation}
Due to the presence of various types of vertices ($n$-body interaction,
 $n=1,2,...$), the counting of the sign and symmetry factor cannot
be simplified as in the usual two-body interaction case.
A third remark is that in general ${\cal H}^0$ and ${\cal H}^1$
conserve spin, and so spin is conserved along the lines, and the total 
spin entering a vertex is the same as the total spin leaving it.

\begin{figure}
\vglue 0.5 truecm
\epsfxsize 8truecm\centerline{\epsfbox{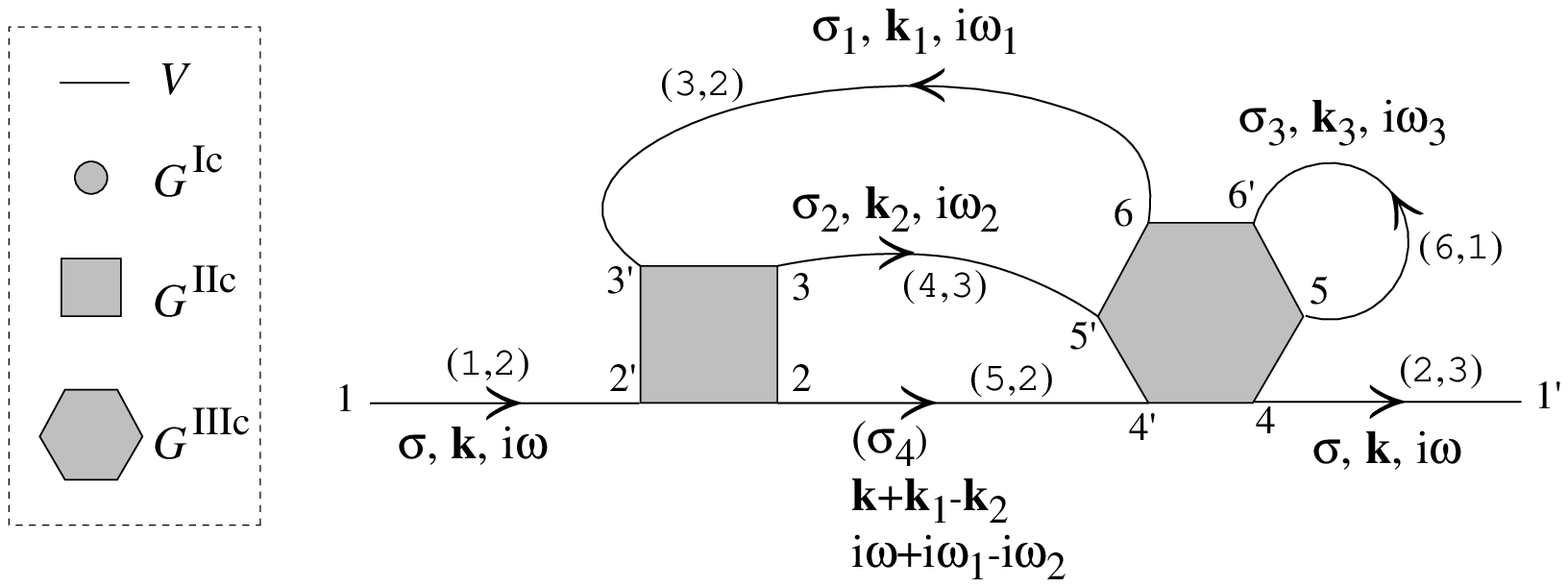}} 
\vglue 0.5 truecm
\caption{
Diagram of order $\vert V\vert^6$ contributing to the Green function
of the auxiliary field ${\cal V}_\si(\vec k,\w)$. 
}
\label{ExempleGComplet}
\end{figure}
\begin{figure}
\vglue 0.5 truecm
\epsfxsize 3.5truecm\centerline{\epsfbox{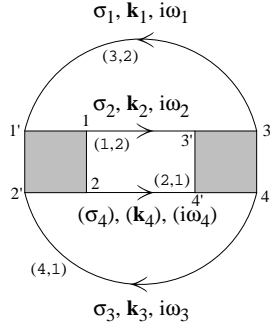}} 
\vglue 0.5 truecm
\caption{
Diagram of order $\vert V\vert^4$ contributing to the thermodynamic
potential. 
}
\label{ExempleOmComplet}
\end{figure}

After a Fourier transform, space and time translation invariance of 
the on-site Hamiltonian $h$ result in the conservation of total momentum 
and total Matsubara frequency at each vertex.
The rules remain the same, except that one can {\it a priori} impose 
the above conservation laws along the lines and at each vertex.
There remain $R_1+...+R_P-R_0-P+1$ independent internal variables 
entering sums of the following type:
\begin{equation}
{1\over \be L^d}\sum_{\si, \vec k,ik_n}...,
\end{equation}
where $L$ is the size of the lattice, $\vec k$ a vector of the reciprocal 
lattice, and $ik_n=i(2n+1)\pi T$ a fermionic Matsubara frequency.
Finally, if the on-site Hamiltonian $h$ is sufficiently simple as to allow 
for an exact computation of the vertices (the $G^{R\rm c}$'s), we can obtain, 
by usual diagrammatic perturbation theory, any correlation function of 
the auxiliary field. 
We still have to derive from them the electron correlation functions: 
this is done in Sect.~\ref{Relations}.
Fig.~\ref{ExempleGComplet} shows an example of a diagram of order 6
contributing to ${\cal V}_{ab}=-\moy{\ps_a\ps^{\star}_b}$.
The algebraic expression corresponding 
to Fig.\ref{ExempleGComplet} is
\begin{eqnarray}
S \sum_{\si_1,\si_2,\si_3}&&{V(\vec k)^2\over (\be L^d)^4}
\sum_{\vec k_1,\vec k_2,\vec k_3\atop \w_1,\w_2,\w_3}
V(\vec k_1)V(\vec k_2)V(\vec k_3)V(\vec k+\vec k_1-\vec k_2) \NN\\
&&\times G^{\rm IIc}_{\si,\si_1;(\si_4),\si_2}(\w,\w_1;\w+\w_1-\w_2,\w_2) \NN\\ 
&&\times G^{\rm IIIc}_{(\si_4),\si_2,\si_3;\si,\si_3,\si_1}
(\w+\w_1-\w_2,\w_2,\w_3;\w,\w_3,\w_1),
\end{eqnarray}
where $S$ is the symmetry and sign factor.
The value of $\si_4$ can always be inferred from that of the other spins.
In the present case, $R_1+R_2=2+3=5$, so the global sign is opposite to that of 
the permutation between the unprimed and the primed indices: $-1$.
The symmetry factor is determined by the numbers between parentheses: the first 
one is the order in which the lines were drawn, and the second one the
number of equivalent possibilities of drawing them. The result is
\begin{equation}
S=-{2\times 3\times 3\times 2\times 1\times 1\over (1!)(1!)(2!)^2(3!)^2}
=-{1\over 4}.
\end{equation}

The rules for the thermodynamic potential are nearly the same. 
From Eq.~(\ref{z2}) one has 
\begin{equation}
\label{ThermoPot}
\Omega=\Omega _0-T\log \int [d\ps ^{\star}d\ps] \e^{\Bra\ps V^{-1}\Ket\ps}
\moy{\e^{\sca\ps\ga + \sca\ga\ps}}_0.
\end{equation}
We know $\Omega_0$ exactly, and the second term is
obtained by the sum of all connected diagrams with no external points,
times $-T$. If we consider the thermodynamic potential per
site, the factor becomes $-1/(\be L^d)$.
When computing the diagrams in momentum space, the factor
$1/(\be L^d)$ is already taken into account by the definition
of the various Fourier transforms.
Fig.~\ref{ExempleOmComplet} shows a diagram contributing to the 
thermodynamic potential at order 4.
The algebraic expression corresponding to Fig.\ref{ExempleOmComplet} is
\begin{eqnarray}
S \sum_{\si_1,\si_2,\si_3}&&{1\over (\be L^d)^3}
\sum_{\vec k_1,\vec k_2,\vec k_3\atop \w_1,\w_2,\w_3}
V(\vec k_1)V(\vec k_2)V(\vec k_3)V(\vec k_1+\vec k_2-\vec k_3)\NN\\
&&\times G^{\rm IIc}_{\si_1,\si_2;\si_3,(\si_4)}(\w_1,\w_2;\w_3,\w_1+\w_2-\w_3)
\NN\\ 
&&\times G^{\rm IIc}_{\si_3,(\si_4);\si_1,\si_2}(\w_3,\w_1+\w_2-\w_3;\w_1,\w_2).
\end{eqnarray}
Here, $R_1+R_2=2+2=4$, so the global sign is opposite 
(see Eq.~(\ref{ThermoPot}) and the following paragraph) to that of
the permutation between the unprimed and the primed indices: $-1$.
The symmetry factor is
\begin{equation}
S=-{2\times 1\times 2\times 1\over (2!)(2!)^2(2!)^2}=-{1\over 8}.
\end{equation}
%

\section{Atomic Correlation Functions}
\label{Atomique}
This appendix gives the one-particle and two-particle correlation functions of
the atomic limit  of the Hubbard model. The chemical potential is $\mu$ and
there is a uniform magnetic field $h$. 
\begin{equation}
{\cal H}_{\rm at}=
Uc_{\up}^{\dag}c_{\dow}^{\dag}c_{\dow}c_{\up}\>
-\mu (c_{\up}^{\dag}c_{\up}+c_{\dow}^{\dag}c_{\dow})
-h(c_{\up}^{\dag}c_{\up}-c_{\dow}^{\dag}c_{\dow}).
\end{equation}
%

\subsection{Green function}
The Green function
\begin{equation}
G_{\si}(\ta_1,\ta_2)=
-\moy{T_{\ta}c_{\si}(\ta_1)c^{\dag}_{\si}(\ta_2)}
\end{equation} 
is straightforward to obtain. Introducing the mean occupation
for each spin
\begin{equation}
n_\pm ={ \e^{(\mu\pm h)\be}+\e^{(2\mu-U)\be} \over 
1+\e^{(\mu +h)\be} +\e^{(\mu - h)\be} + \e^{(2\mu-U)\be} },
\end{equation} 
the Fourier transform of the Green function reads:
\begin{equation}
\label{Gat}
G_{\pm}(\w)={1-n_{\mp}\over\w+\mu+\pm h}+{n_{\mp}\over \w+\mu+\pm h-U}
\qquad (\si=\pm)\;.
\end{equation}
%

\subsection{Two-particle function}

We start from the definition of $G^{\rm II}$ in imaginary time 
(by enclosing $\si_4$ between parentheses, we mean that its value 
is to be inferred from that of the other three spins):
\begin{equation}
G^{\rm II}_{\si_1\si_2,(\si_4)\si_3}(\ta_1,\ta_2;\ta_4,\ta_3)=
\moy{T_{\ta}c_{\si_1}(\ta_1)c_{\si_2}(\ta_2)
c^{\dag}_{\si_3}(\ta_3)c^{\dag}_{(\si_4)}(\ta_4)}
\end{equation} 
and compute it explicitly for any time ordering, using $\e^{-(H-\mu N)\tau}$ as
evolution operator and inserting complete sets of states as necessary.
We then calculate its Fourier transform,  defined by the following equation:
\begin{eqnarray}
\int_0^\be d\ta_1 ... d\ta_4\> 
G^{\rm II}_{\si_1\si_2,(\si_4)\si_3}(\ta_1,\ta_2;\ta_4,\ta_3)
\>\e^{\w_1\ta_1+\w_2\ta_2-\w_3\ta_3-\w_4\ta_4}= \\
\be\de(\w_1+\w_2-\w_3-\w_4)\>
G^{\rm II}_{\si_1\si_2,(\si_4)\si_3}(\w_1,\w_2;(\w_4),\w_3).
\end{eqnarray} 
Again, $\w_4=\w_1+\w_2-\w_3$.
After introducing the following short-hand notation
\begin{eqnarray}
z=1+\e^{(\mu +h)\be}+\e^{(\mu - h)\be}+\e^{(2\mu-U)\be}, \\
\x_j^\pm=\w_j+\mu\pm h, \\
\xb_j^\pm=\w_j+\mu\pm h-U,
\end{eqnarray}
the connected correlation functions read:
\begin{equation}
\label{GIIpara}
G^{\rm IIc}_{\up\up,\up\up}(\w_1\w_2,(\w_4)\w_3)=
{\be U^2 n_-(1- n_-)[\de (\w_2-\w_3 )-\de (\w_1-\w_3 )]\over
\x_1^+\xb_1^+ \x_2^+\xb_2^+ },
\end{equation}
and
\begin{mathletters}
\begin{eqnarray}
\label{GIIanti}
G^{\rm IIc}_{\dow \up,\dow \up }&&(\w_1\w_2,(\w_4)\w_3) = \\
\label{dope}
&&{n_+ +n_- -1\over \w_1+\w_2+2\mu -U}
\l( {1\over\xb_1^-}+{1\over\xb_2^+} \r)
\l( {1\over\xb_3^+}+{1\over\xb_4^-} \r) \\
\label{mag}
&+&{n_+ -n_- \over \w_1-\w_3-2h}\l( {1\over \x_1^-}-{1\over\xb_3^+} \r)
\l( {1\over \x_4^-}-{1\over\xb_2^+}\r) \\
\label{con}
&+&{\be U^2\de (\w_2-\w_3)(\e^{(2\mu-U)\be}-\e^{2\mu\be})\over z^2}
{1\over \x_1^-\xb_1^- \x_2^+\xb_2^+} \\
\label{norm1}
&+&{n_+-1\over \x_1^-\xb_3^+ \x_4^-}+
{1-n_+\over \x_1^-\xb_2^+ \xb_3^+}+
{1-n_-\over \xb_1^-\x_2^+ \xb_3^+}+
{n_--1\over \x_2^+\xb_3^+ \x_4^-}+
{1-n_-\over \x_1^- \x_2^+ \x_4^-}+
{1-n_-\over \x_1^- \x_2^+ \x_3^+} \\
\label{norm2}
&+&{1-n_-\over \xb_1^- \x_3^+ \xb_4^-}+
{n_--1\over \xb_1^- \x_2^+ \x_3^+}+
{1-n_-\over \xb_2^+ \x_3^+ \xb_4^-}+
{n_--1\over \x_1^-\xb_2^+ \x_3^+}+
{-n_+\over \xb_1^-\xb_2^+ \xb_4^-}+
{-n_+\over \xb_1^-\xb_2^+ \xb_3^+}.
\end{eqnarray}
\end{mathletters}
All values of $G^{\rm II c}_{\si_1\si_2,(\si_4)\si_3}(\w_1\w_2,(\w_4)\w_3)$ can
be deduced from Eqs.~(\ref{GIIpara}) to (\ref{norm2}) above,  the antisymmetry
with respect to the exchange of two incoming or outgoing variables, and the
possibility to exchange $\up$ and $\dow$ by doing $h\to -h$.

Apart from the regular terms of Lines (\ref{norm1}) and (\ref{norm2}), 
$G^{\rm II c}$ contains only singular or possibly singular terms.  Indeed,
Line~(\ref{con}) has $\de (\w_2-\w_3)$ in factor, Line~(\ref{dope}) becomes
proportional to  $\de (\w_1+\w_2)$ in the half-filling limit,  and
Line~(\ref{mag}) becomes proportional to $\de (\w_1-\w_3)$  in the zero
magnetic field limit.  These special cases have to be properly taken into
account while computing $G^{\rm II c}$ directly at half-filling  and zero
magnetic field.

\begin{figure}
\vglue 0.5 truecm
\epsfxsize 3truecm\centerline{\epsfbox{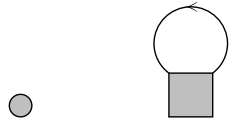}} 
\vglue 0.5 truecm
\caption{
Diagrams contributing to $\Ga$ up to order $t_0$.
}
\label{DiagTest}
\end{figure}
\begin{figure}
\vglue 0.5 truecm
\epsfxsize 8truecm\centerline{\epsfbox{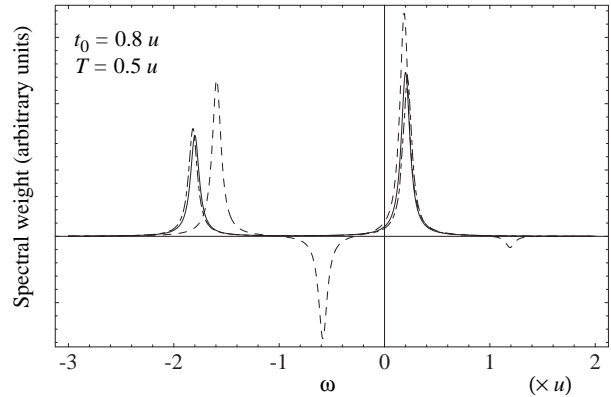}} 
\vglue 0.5 truecm
\caption{
Spectral function of the atomic model with chemical potential $\mu=1.8u$
at temperature $T=0.5u$. Comparison between the exact result (solid), 
the raw approximation (dashed), and the continued fraction (dot-dashed).
Whereas the dashed curve is neither normalized, nor even positive, the
difference between the other two curves is hardly noticeable.
}
\label{Test}
\end{figure}
\begin{figure}
\vglue 0.5 truecm
\epsfxsize 8truecm\centerline{\epsfbox{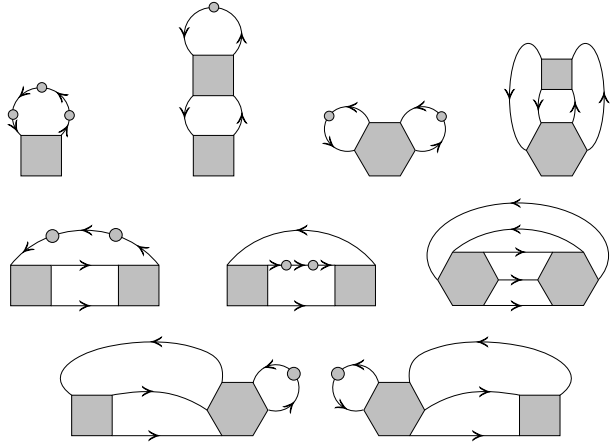}} 
\vglue 0.5 truecm
\caption{
Diagram contributing to $\Ga$ at fourth order (top row)  and fifth order
(bottom two rows). }
\label{Ordres45}
\end{figure}

\section{A Practical Test of The Method}
\label{TestProc}
In this section, we test the diagrammatic theory and the continued
fraction representation on an exactly solvable simple model.
We consider the atomic limit itself, with a slight doping given by a
chemical potential $\mu=u+t_0$.
On the one hand, we know the exact Green function from Eq.~(\ref{Gat}):
\begin{equation}
G_{\rm ex}(\w)={1-n\over\w+u+t_0}+{n\over \w-u+t_0},
\end{equation} 
with
\begin{equation}
n={ \e^{(u+t_0)\be}+\e^{2t_0\be} \over 
1+2\e^{(u+t_0) \be} +\e^{2t_0\be} }.
\end{equation} 
On the other hand, if we consider the chemical potential shift as a zero-range
hopping perturbation, we can compute $\Ga$ to first order in powers of $t_0$
with the help of the diagrams depicted in Fig.~\ref{DiagTest}, and obtain
\begin{equation}
\Ga^{(1)}(\w)={\w\over\wp^2-u^2}+{-t_0u^2\over(\wp^2-u^2)^2}+
{t_0\be u \nf (u)\over\wp^2-u^2},
\end{equation}
where $\nf$ is the Fermi occupation factor. It is straightforward to check that
the exact Green function and the  ``raw'' approximation given by the above
value of $\Ga$ 
\begin{equation}
G^{(1)}_{\rm raw}(\w)={1\over 
\Ga^{(1)}(\w)^{-1} +t_0}.
\end{equation}
coincide  up to order $t_0$ included. But this expansion of $\Ga$,
although undoubtedly correct, leads to the general causality problem described
in  Sect.~\ref{FracTrick}. On the other hand, the reconstructed continued
fraction is
\begin{equation}
G_J^{(1)}(\w)={1\over{\w+(1-\be u\nf (u))t_0-
{\DS u^2\over\DS \w+(1+\be u\nf (u))t_0}}}.
\end{equation}
Fig.\ref{Test} shows a comparison between the spectral weight of the exact
solution,  the raw approximation, and the final continued fraction. One can see
that $G_J^{(1)}(\w)$ yields an extremely good spectral function (which remains
true at any temperature and for a very wide range  of values of $t_0$).

Nonetheless, our enthusiasm has to be tempered by the following remarks.
First, the exact solution being a two-pole rational function, it is not
surprising that a finite continued fraction is able to mimic it satisfactorily.
Secondly, going to low temperature does no harm in this case, but the two-site
problem, described in Sect.~\ref{two-site},  shows that this is not true in
general.

\section{Higher loops and the two-site Problem}
\label{HODS}

\subsection{Order $t^4$ and order $t^5$}
\label{HO}

Fig.~\ref{Ordres45} shows the diagrams contributing to fourth and fifth order.
The diagrams of Fig.~\ref{Ordres45} lead to the following expression for $\Ga$:
\begin{eqnarray}
\Ga^{(4)}(\vec k,\w)&=&{\frac{\DS \w}{\DS {\wp^2-1}}} + 
{\frac{\DS 6d{t^2}\w}
 {\DS {{\big( {\wp^2} -1\big) }^3}}} + 
 6{{c(\vec k)}}{t^3}
 \Bigg( {\frac{\DS {\be }\l( -1 + {e^{\be }} \r) }
 { \DS 2\l( 1 + {e^{\be }} \r) 
 {{\big( {\wp^2} -1\big) }^2}}}
 {\frac{\DS 2{\wp^2}-1}
 { \DS {{\big( {\wp^2}-1 \big) }^4}}} \Bigg) \NN\\
&+{t^4}\Bigg\{ & {\frac{\DS 3{{\be }^2}d
 \l( -1 + 2d \r) {e^{\be }}{(\w)}}{\DS
 {{\l( 1 + {e^{\be }} \r) }^2}
 {{\l( \w-1 \r) }^3}
 {{\l( \w+1 \r) }^3}}} + 
{\frac{\DS 3{\be }d\l( -1 + {e^{\be }} \r) 
 \w}{\DS
 {{\l( 1 + {e^{\be }} \r) }^3}
 {{\l( \w-1 \r) }^4}
 {{\l( \w+1 \r) }^4}}} \NN\\
&\times & \l( -2 - 13{e^{\be }} + 18d{e^{\be }} - 
 2{e^{2{\be }}} + 2{\wp^2} + 
 {e^{\be }}{\wp^2} + 6d{e^{\be }}{\wp^2} + 
 2{e^{2{\be }}}{\wp^2} \r) \NN\\
&+& {\frac 
	{\DS 3d{\w}}
	{\DS 2{{\l( 1 + {e^{\be }} \r) }^2}
 \l( \w-3 \r) 
 {{\l( \w -1\r) }^5}
 {{\l( \w+1 \r) }^5}
 \l( \w +3\r) }} \NN\\ 
&\times \Big[ & (247 - 278d + 914{e^{\be }} - 
 1396d{e^{\be }} + 247{e^{2{\be }}} -
	 278d{e^{2{\be }}} ) \NN\\
&+ & (15 - 378d + 450{e^{\be }} -1596d{e^{\be }} 
 + 15{e^{2{\be }}} - 378d{e^{2{\be }}}) {\wp^2} \NN\\
&+ & (- 3 + 42d - 90{e^{\be }} + 252d{e^{\be }} 
 - 3{e^{2{\be }}} + 42d{e^{2{\be }}}){\wp^4} \NN\\
&+ & (- 3 + 6d + 6{e^{\be }} - 12d{e^{\be }} - 3{e^{2{\be }}}
 + 6d{e^{2{\be }}}) {\wp^6}\Big]\Bigg\} ,
\end{eqnarray}
where we have set $u=1$.
The fifth order and the corresponding continued fraction 
are too lengthy to be presented here, but are available on the
internet.\cite{Pai98}
\begin{figure}
\vglue 0.5 truecm
\epsfxsize 7truecm\centerline{\epsfbox{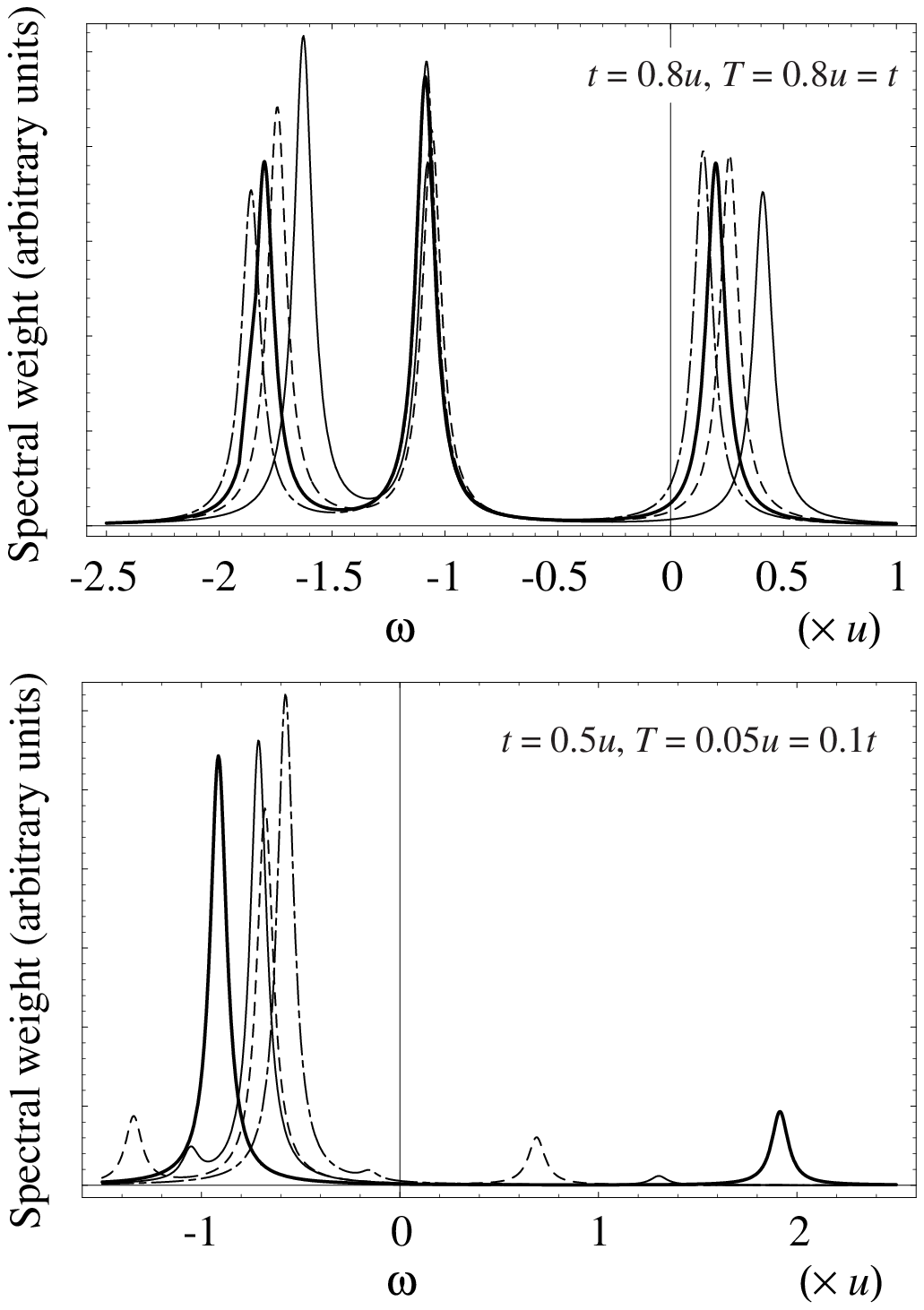}} 
\vglue 0.5 truecm
\caption{
Above: Spectral function $A(k=0,\om )$ of the two-site Hubbard model with hopping
term $t=0.8u$ at temperature $T=t=0.8u$. Comparison between the exact result
(bold) and the continued fraction at third (solid), fourth (dashed) and fifth
(dot-dashed) order. In this region of parameter space, the approximation
improves with increasing order. Below: The same, but with $t=0.5u$ at temperature
$T=0.1t=0.05u$. In this region of parameter space, the third-order
approximation is the best one.}
\label{DeuxSites}
\end{figure}

\subsection{The two-site problem}
\label{two-site}
The exactly solvable two-site problem provides a good 
testing ground for the higher-order results just obtained. In this 
subsection the parameter $t$ is replaced by $t/2$ in order to account
for periodic boundary conditions on a lattice with only two sites (say site 1
and site 2).  The analytic expressions ${\cal G}_{11}(\w)$ and ${\cal
G}_{12}(\w)$ of the Green function are available.
Replacing the momentum integrations by sums over the two 
values $k=0$ and $k=\pi$, the expansion of ${\cal G}_{11}$ 
up to order $t^5$ has been computed in the usual way (similar to what was
done in the previous subsection), and checked to be the same 
as the one obtained from the exact solution.
Furthermore, it leads to a continued fraction having only four floors
instead of eight, as it should since the exact solution has four poles.
It is then possible to investigate the validity of the various
approximations throughout the $(T,t)$ plane.
Empirically, we were led to following conclusion: for $T>t$ the 
approximation manifestly improves with the order, and for $T\leq t$ the
approximation deteriorates much faster with increasing order.
Two examples illustrating this behavior are shown in 
Fig.~\ref{DeuxSites}.


\begin{references}
\bibitem{note-Boies}
The $1/(R!)^2$ prefactor corrects Eq.~(5) of Ref.\onlinecite{Boi95}, where
  $1/(n!)^2$ should have appeared.
\end{references}

\end{document}